\documentclass[10pt,a4paper]{article}
\usepackage[a4paper]{geometry}

\usepackage{amssymb,latexsym,amsmath,amsfonts,amsthm}
\usepackage{graphicx}
\usepackage{overpic}

    \DeclareMathOperator*{\supp}{supp}
    \DeclareMathOperator*{\diag}{diag}
    \DeclareMathOperator{\repart}{Re}
    \DeclareMathOperator{\impart}{Im}
    \renewcommand{\Re}{\repart}
    \renewcommand{\Im}{\impart}

    \DeclareMathOperator*{\Tr}{Tr}
    \newcommand{\Res}{\operatorname{Res}}
    \newcommand{\lam}{\lambda}

    \newcommand{\sign}{{\operatorname{sgn}\,}}

    \newtheorem{theorem}{Theorem}[section]
    \newtheorem{lemma}[theorem]{Lemma}
    
    \newtheorem{proposition}[theorem]{Proposition}
    
    \newtheorem{Definition}[theorem]{Definition}
    
    \newtheorem{Remark}[theorem]{Remark}
    
    \newtheorem{Example}[theorem]{Example}
    \newenvironment{example}{\begin{Example}\rm}{\end{Example}}

\numberwithin{equation}{section}

\title{Random matrix model with external source and a constrained vector equilibrium problem}

\author{Pavel Bleher\footnotemark[1]\ , Steven Delvaux\footnotemark[2]\ , and Arno
B.J. Kuijlaars\footnotemark[2]}

\date{\today}

\begin{document}

\maketitle
\renewcommand{\thefootnote}{\fnsymbol{footnote}}
\footnotetext[1]{Department of Mathematical Sciences, Indiana University-Purdue
University Indianapolis, 402 N. Blackford St., Indianapolis, IN 46202, U.S.A.
email: bleher\symbol{'100}math.iupui.edu.} \footnotetext[2]{Department of
Mathematics, Katholieke Universiteit Leuven, Celestijnenlaan 200B, B-3001
Leuven, Belgium. email:
\{steven.delvaux,arno.kuijlaars\}\symbol{'100}wis.kuleuven.be.}

\begin{abstract} We consider the random matrix model with external source, in case where the
potential $V(x)$ is an even polynomial and the external source has two
eigenvalues $\pm a$ of equal multiplicity. We show that the limiting
mean eigenvalue distribution of this model can be characterized as the
first component of a pair of measures $(\mu_1,\mu_2)$
that solve a constrained vector equilibrium problem. The proof is based
on the steepest descent analysis of the associated Riemann-Hilbert problem
for multiple orthogonal polynomials.

We illustrate our results in detail for the case of a quartic double well
potential $V(x) = \frac{1}{4} x^4 - \frac{t}{2} x^2$. We are able
to determine the precise location of the phase transitions in the $ta$-plane,
where either the constraint becomes active, or the two intervals in
the support come together (or both).
\end{abstract}

\section{Introduction}
\label{section:intro}

The random matrix model with external source is the probability measure
\begin{equation} \label{eq:sourcemodel}
    \frac{1}{Z_n} \exp \left( - n \Tr (V(M) - AM) \right) dM
    \end{equation}
defined on the space of $n \times n$ Hermitian matrices $M$.
Here $A$ is a given Hermitian matrix (the external source),
 $V: \mathbb R \to \mathbb R$ is a function with sufficient
increase at infinity (the potential), and $Z_n$
is a normalization constant.

The model \eqref{eq:sourcemodel} was first studied by Br\'ezin and Hikami
\cite{BH1,BH2} and P.~Zinn-Justin \cite{ZJ1,ZJ2} who showed that the eigenvalue
correlations are determinantal. In \cite{BK1} it was observed that the
correlation kernel can be expressed in terms of multiple orthogonal
polynomials. Due to the Riemann-Hilbert problem for multiple orthogonal
polynomials \cite{VAGK} this opened up a new way for asymptotic analysis. For
the quadratic case
\begin{equation} \label{eq:quadraticV}
    V(x) = \frac{1}{2} x^2
    \end{equation}
with external source
\begin{equation} \label{eq:sourceA}
    A = \diag(\underbrace{a, \ldots, a}_{n/2 \text{ times}},
    \underbrace{-a, \ldots, -a}_{n/2 \text{ times}})
    \end{equation}
with two eigenvalues $\pm a$ of equal multiplicity (thus $n$ is even), this was
done in great detail in the three papers \cite{ABK,BK2,BK3}.

The quadratic case is of special interest because it has an equivalent
formulation in terms of non-intersecting Brownian motions that start at one
value and end at certain prescribed values \cite{ABK} which is a variation on
Dyson's Brownian motion \cite{Dyson}.
The quadratic model with external source \eqref{eq:sourceA}
exhibits a phase transition, since for small $a > 0$ the eigenvalues
accumulate on one interval
while for larger $a$ the eigenvalues accumulate on two disjoint intervals.
At the critical value of $a$ the local eigenvalue correlations are
given in terms of Pearcey integrals
\cite{AvM1,BK3,OR,TW1}.

In this paper we study the external source model \eqref{eq:sourcemodel} with a
more general potential $V$. We assume that $V$ is an even polynomial
\begin{equation} \label{eq:polynomialV}
    V(x) = \sum_{j=1}^{d} v_j x^{2j}, \qquad v_d > 0
    \end{equation}
of degree $2d$.

For $a=0$ the external source model \eqref{eq:sourcemodel} reduces to the usual
unitary matrix model
\begin{equation} \label{eq:unitarymodel}
    \frac{1}{Z_n} \exp\left( - n \Tr V(M) \right),
    \end{equation}
which is one of the most studied models in random matrix theory in both
mathematics and physics, see e.g.\ \cite{BI2,BI3,CK,CV,DKMVZ1} for rigorous study
using the Riemann-Hilbert approach. A basic fact is that for $n\to\infty$,
the limiting mean eigenvalue
distribution of the matrix $M$ in \eqref{eq:unitarymodel} minimizes the energy
functional
\begin{equation} \label{eq:energyfunctional:scalar}
    E(\mu) = \iint \log \frac{1}{|x-y|} d\mu(x) d\mu(y)
    + \int V(x) d\mu(x)
    \end{equation}
over all probability measures $\mu$ on $\mathbb R$.

It is an open problem to find an analogue for the equilibrium problem
\eqref{eq:energyfunctional:scalar} in the general context of the random matrix
model with external source. This paper contains a first result in this
direction. We consider the external source model \eqref{eq:sourcemodel} in case
where the potential $V$ is an even polynomial \eqref{eq:polynomialV}. The
external source $A$ is again given by \eqref{eq:sourceA} with two eigenvalues
$\pm a$ of equal multiplicity. We show that under these assumptions, the
limiting mean eigenvalue distribution of the matrix $M$ in
\eqref{eq:sourcemodel} exists, and that it arises as the first component of a
pair of measures $(\mu_1,\mu_2)$ solving a certain vector equilibrium problem,
see Section~\ref{section:statementresults}.

We will illustrate our results in detail for a particular case of a non-convex
potential, namely the quartic double well potential
\begin{equation} \label{eq:quarticV}
    V(x) = \frac{1}{4} x^4 - \frac{t}{2} x^2, \qquad t > 0.
    \end{equation}
For the quartic model \eqref{eq:unitarymodel}, \eqref{eq:quarticV} (without
external source) it is known that the eigenvalues accumulate on either one or
two intervals. The local eigenvalue correlations for the critical
value of $t = t_{cr} = 2$ are given in terms of $\Psi$-functions associated
with the Hastings-McLeod solution of the Painlev\'e II equation \cite{BI2,CK}.

So in the quartic model with external source there exist at least two
mechanisms by which a transition from one to two intervals can occur: namely a
Pearcey transition and a Painlev\'e II transition. It will be one of the
outcomes of the present paper that we can determine precisely the location
of the phase transitions in the $ta$-plane.

\section{Statement of results}
\label{section:statementresults}

\subsection{Equilibrium problem}
\label{subsection:equilibriumproblem}

The main ingredient in our analysis is a new vector equilibrium problem
associated with the random matrix model \eqref{eq:sourcemodel} with external
source. We emphasize that it only applies in the setting we are considering,
namely an even polynomial potential $V$ as in \eqref{eq:polynomialV} and an
external source \eqref{eq:sourceA} with two eigenvalues of equal multiplicity.
This setting gives a symmetry with respect to the origin, which we use in an
essential way.

The equilibrium problem is as follows. We minimize the energy
functional
\begin{multline} \label{eq:energyfunctional}
    E(\mu_1,\mu_2) = \iint \log \frac{1}{|x-y|} d\mu_1(x) d\mu_1(y)
    +\iint \log \frac{1}{|x-y|} d\mu_2(x) d\mu_2(y) \\
    - \iint \log \frac{1}{|x-y|} d\mu_1(x) d\mu_2(y)
    + \int \left(V(x) - a |x|\right) d\mu_1(x)
    \end{multline}
with respect to all pairs of measures $(\mu_1,\mu_2)$ satisfying
\begin{itemize}
\item $\mu_1$ and $\mu_2$ have finite logarithmic energy,
\item $\mu_1$ is a measure on $\mathbb R$ with total mass $1$,
\item $\mu_2$ is a measure on $i \mathbb R$ with total mass $1/2$
that satisfies the constraint
\begin{equation} \label{eq:upperconstraint}
    \mu_2 \leq \sigma
    \end{equation}
where $\sigma$ is the measure on $i\mathbb R$ with constant density
\begin{equation} \label{eq:constraintsigma}
    \frac{d\sigma}{|dz|} = \frac{a}{\pi}, \qquad z \in i \mathbb R.
    \end{equation}
\end{itemize}

Standard references on potential theory in the complex plane are
\cite{Dei,NS,SaffTotik}.

The equilibrium problem \eqref{eq:energyfunctional} has both an external field
$V(x) - a|x|$ acting on $\mu_1$, and an upper constraint $\sigma$ acting on
$\mu_2$. The interaction between $\mu_1$ and $\mu_2$ is of Nikishin type
\cite{Ku2}. This type of vector equilibrium problem also appeared recently in a
model of non-intersecting squared Bessel paths \cite{KMFW} and in the
two-matrix model with quartic potential \cite{DuK}.

Our first result concerns the structure of the minimizer of the equilibrium
problem.

\begin{theorem} \label{theorem:minimizer}
There is a unique minimizer $(\mu_1,\mu_2)$ which satisfies
\begin{enumerate}
\item[\rm (a)] The support of $\mu_1$ is bounded and consists of a finite union
of intervals
\begin{equation}\label{def:N}
    \supp(\mu_1) = \bigcup_{j=1}^N [a_j, b_j].
\end{equation}
The measure $\mu_1$ is absolutely continuous with density
\begin{equation}\label{density:mu1}
    \frac{d\mu_1(x)}{dx} = h(x) \sqrt{\prod_{j=1}^{N}(b_j-x)(x-a_j)},
    \qquad x \in \bigcup_{j=1}^{N}[a_j,b_j],
\end{equation}
where $h$ is a nonnegative function on $\supp(\mu_1) = \bigcup_{j=1}^{N} [a_j,b_j]$
that is real analytic, except possibly at zero.
\item[\rm (b)] The support of $\mu_2$ is the full imaginary axis and there
exists $c \geq 0$ such that
\begin{equation}\label{def:c}
     \supp(\sigma-\mu_2) = (-i \infty, -ic] \cup [ic, i\infty).
\end{equation}
We have that $c=0$ if and only if
\begin{equation} \label{condition:CaseI}
    \int  \frac{d\mu_1(s)}{|s|} \leq 2 a
    \end{equation}
and in that case, $\mu_2$ has the density
\begin{equation}\label{density:mu2a}
    \frac{d\mu_2(z)}{|dz|} = \frac{1}{2\pi}\int \frac{|s|}{|z|^2+s^2}\,d\mu_1(s),
    \qquad z \in i \mathbb R.
\end{equation}
If \eqref{condition:CaseI} is not satisfied then $c > 0$ is determined
by the condition
\begin{equation} \label{condition:CaseIIorIII}
    \int \frac{d\mu_1(s)}{\sqrt{s^2 + c^2}} = 2a,
\end{equation}
and in that case
\begin{equation}\label{density:mu2b}
    \frac{d\mu_2(z)}{|dz|} = \frac{a}{\pi}, \qquad \text{if } z \in [-ic,ic],
\end{equation}
and
\begin{align} \label{density:mu2c}
\frac{d\mu_2(z)}{|dz|} &= \frac{a}{\pi} - \frac{1}{2\pi}\int
\frac{|z|\sqrt{|z|^2-c^2}}{(|z|^2+s^2)\sqrt{s^2+c^2}}\,d\mu_1(s) \\
\label{density:mu2d}
    &= \frac{1}{2\pi}\int
\frac{|z|^2+s^2-|z|\sqrt{|z|^2-c^2}}{(|z|^2+s^2)\sqrt{s^2+c^2}}\,d\mu_1(s),
\end{align}
if $z\in(-i\infty,-ic]\cup[ic,i\infty)$.
\item[\rm (c)] Both $\mu_1$ and $\mu_2$ are symmetric with respect
to the origin.
\end{enumerate}
\end{theorem}
Theorem~\ref{theorem:minimizer} will be proved in
Sections~\ref{subsection:prooftheoremminimizer1} and
\ref{subsection:prooftheoremminimizer2}. Note that
\eqref{def:N}--\eqref{density:mu2a} are similar to statements proved in
\cite{DuK}, while \eqref{condition:CaseIIorIII} and
\eqref{density:mu2c}--\eqref{density:mu2d} have apparently not been stated
before.

\subsection{Variational conditions}
\label{subsection:variational}

The minimizer $(\mu_1,\mu_2)$ to the equilibrium problem in
Section~\ref{subsection:equilibriumproblem} is characterized by the following
Euler-Lagrange variational conditions. We write
\[ U^{\mu}(x) = \int \log \frac{1}{|x-y|} d\mu(y) \]
for the logarithmic potential of a measure $\mu$.

\begin{proposition}\label{prop:variational}
The measures $\mu_1$ and $\mu_2$ satisfy for some constant $\ell \in \mathbb
R$,
\begin{align} \label{eq:varcondition1}
    2 U^{\mu_1}(x) & = U^{\mu_2}(x) - V(x) + a |x| - \ell,
        \qquad x \in \supp(\mu_1), \\
        \label{eq:varcondition2}
    2 U^{\mu_1}(x) & \geq U^{\mu_2}(x) - V(x) + a |x| - \ell,
        \qquad x \in \mathbb R \setminus \supp(\mu_1), \\
        \label{eq:varcondition3}
    2 U^{\mu_2}(x) & = U^{\mu_1}(x),
        \qquad x \in \supp(\sigma - \mu_2), \\
        \label{eq:varcondition4}
    2 U^{\mu_2}(x) & < U^{\mu_1}(x),
        \qquad x \in i \mathbb R \setminus \supp(\sigma - \mu_2).
\end{align}
\end{proposition}

These relations follow directly from the variational conditions of the
equilibrium problem.

\subsection{Regular and singular cases}

We say that $\mu_1$ is regular if in \eqref{density:mu1} we have $h(x)
> 0$ on $\supp(\mu_1)=\bigcup_{j=1}^{N}[a_j,b_j]$, and if the variational inequality
\eqref{eq:varcondition2} is strict for every $x \in \mathbb R \setminus
\supp(\mu_1)$. Otherwise $\mu_1$ is called singular \cite{DKMVZ1,DKMVZ2}.

The measure $\mu_2$ has a density on $i \mathbb R$ which is bounded  by
$\frac{a}{\pi}$. We say that $\mu_2$ is singular if equality in this
restriction is attained at $z=0$ and at no other point of $i\mathbb R$. In all
other cases (in particular if $c > 0$), the measure $\mu_2$ is called regular.

For the analysis in this paper, we will assume that both $\mu_1$ and $\mu_2$
are regular. The following lemma follows immediately from this assumption and
is stated only for further reference.

\begin{lemma}\label{prop:squareroot} The measures
$\mu_1$ and $\sigma-\mu_2$ satisfy the following square root behavior near
their endpoints $a_j$, $b_j$ and $\pm ic$:
\begin{enumerate}
\item[\rm (a)] If the measure $\mu_1$ is regular then it has a density of the form \eqref{density:mu1}
with $h$ strictly positive on $\bigcup_{j=1}^{N}[a_j,b_j]$.
\item[\rm (b)]
If $c>0$ then the measure $\sigma-\mu_2$ has a density of the form
\[ \frac{a}{\pi}-\frac{d\mu_2(z)}{|dz|} = k(z) \sqrt{|z|^2-c^2},\]
    where $k$ is an analytic, strictly positive function on $(-i \infty, -ic] \cup [ic, i \infty)$.
\end{enumerate}
\end{lemma}

Lemma \ref{prop:squareroot}(a) follows immediately from the definition of
$\mu_1$ being regular. Lemma \ref{prop:squareroot}(b) follows from
\eqref{density:mu2c}.

\subsection{Limiting eigenvalue distribution}

Our main result deals with the global distribution of eigenvalues as $n\to
\infty$.
\begin{theorem}\label{theorem:main}
Let $V$ be an even polynomial, and let $A$ be a diagonal matrix with two
eigenvalues $\pm a$ of equal multiplicity. Let $(\mu_1,\mu_2)$ be the solution
of the equilibrium problem in Section~\ref{subsection:equilibriumproblem}, and
assume that both $\mu_1$ and $\mu_2$ are regular in the sense explained above.
Then the mean eigenvalue
distribution of a matrix $M$ from the random matrix model
\[ \frac{1}{Z_n} \exp \left( - n \Tr (V(M) - AM) \right) dM \]
has the limit $\mu_1$ as $n \to \infty$.
\end{theorem}

We strongly expect that the conclusion of Theorem~\ref{theorem:main} remains
valid in the case where $\mu_1$ and/or $\mu_2$ is singular.

Theorem~\ref{theorem:main} will be proved in
Section~\ref{subsection:proofmaintheorem}.

\subsection{About the proof} \label{section:aboutproof}

The proof of Theorem~\ref{theorem:main} is based on the Riemann-Hilbert problem
for multiple orthogonal polynomials and its connection with the external source
model \eqref{eq:sourcemodel}.

The multiple orthogonal polynomials $P_n(x)$ in question are orthogonal with
respect to the weights
\begin{equation} \label{eq:w}
 w_1(x) = e^{-n (V(x) - ax)}, \qquad w_2(x) = e^{-n(V(x) + ax)}.
\end{equation}
More precisely, $P_n(x)$ is a monic polynomial of degree $n$ that is
characterized by the multiple orthogonality conditions (we assume $n$ is even)
\begin{equation} \label{eq:typeIIMOP}
    \int_{-\infty}^{\infty}
    P_n(x) w_j(x) x^k \, dx = 0,
        \qquad j=1,2, \quad k=0, \ldots, n/2-1.
\end{equation}
The polynomial $P_n(x)$ is also the average characteristic polynomial
\begin{equation} \label{eq:averagecharpoly}
    P_n(z) = \mathbb E \left[ \det(z I_n - M ) \right],
\end{equation}
where the expectation $\mathbb E$ is taken with respect to the model
\eqref{eq:sourcemodel}, see \cite{BK1}.

The Riemann-Hilbert problem (RH problem) for $P_n(x)$ is as follows. We look
for an analytic matrix-valued function $Y : \mathbb C  \setminus \mathbb R \to
\mathbb C^{3 \times 3}$ with jump properties
\begin{equation} \label{eq:Yjump}
    Y_+(x) = Y_-(x)
    \begin{pmatrix} 1 & w_1(x) & w_2(x) \\ 0 & 1 & 0 \\ 0 & 0 & 1 \end{pmatrix},
    \qquad x \in \mathbb R,
\end{equation}
and asymptotic condition
\begin{equation} \label{eq:Yatinfty}
    Y(z) = \left( I + O(1/z)\right)
    \begin{pmatrix} z^n & 0 & 0 \\ 0 & z^{-n/2} & 0 \\ 0 & 0 & z^{-n/2} \end{pmatrix}
    \qquad \textrm{as } z \to \infty.
    \end{equation}

Here and below we use the following standard notation. If $\Gamma$ is an
oriented contour in the complex plane, then the side that is on the left
(right) when traversing $\Gamma$ according to its orientation is called the
$+$-side ($-$-side), and for any $z\in\Gamma$ we use
$Y_+(z)$ ($Y_-(z)$) to denote the limiting values of $Y(z)$ along
the $+$-side ($-$-side) of $\Gamma$.
In \eqref{eq:Yjump} we have the contour $\Gamma=\mathbb R$ oriented from left to right.

The RH problem \eqref{eq:Yjump}--\eqref{eq:Yatinfty} has a unique solution. The
$(1,1)$-entry of $Y(z)$ is the multiple orthogonal polynomial $P_n(z)$
characterized by \eqref{eq:typeIIMOP}.

It is known that the eigenvalues of the random matrix model with external
source \eqref{eq:sourceA} form a determinantal point process with correlation
kernel \cite{BK1}
\begin{equation} \label{eq:KninY}
    K_n(x,y) = \frac{1}{2\pi i(x-y)}
    \begin{pmatrix} 0 & w_1(y) & w_2(y) \end{pmatrix}
    Y_+(y)^{-1} Y_+(x) \begin{pmatrix} 1 \\ 0 \\ 0 \end{pmatrix}.
    \end{equation}
Theorem~\ref{theorem:main} then comes down to the following statement about the
limiting behavior of the kernels $K_n$:
\begin{equation} \label{eq:Knlimit}
    \lim_{n \to \infty} \frac{1}{n} K_n(x,x) = \frac{d\mu_1(x)}{dx},\qquad x\in\mathbb R.
\end{equation}
We will establish \eqref{eq:Knlimit} in
Section~\ref{subsection:proofmaintheorem}, thereby proving
Theorem~\ref{theorem:main}.

From the RH analysis it is possible to obtain universality results for the
local eigenvalue correlations as well. In the regular cases, this leads to the
usual sine kernel in the bulk and Airy kernel at the edge points of the
spectrum. We will not discuss this any further and refer to the papers
\cite{ABK,BK2,BK3,DKMVZ1,DKMVZ2,DuK}, among others, for a detailed analysis in
a similar context.

\subsection{Organization of the paper}

The rest of the paper is organized as follows. In Section~\ref{section:equil}
we discuss the structure of the equilibrium measures and we prove
Theorem~\ref{theorem:minimizer}. In Section~\ref{section:riemannsurface} we
introduce the Riemann surface built from the solution of the equilibrium
problem. Section~\ref{section:steepestdescent} contains the steepest descent
analysis of the RH problem for $Y(z)$, leading to the proof of
Theorem~\ref{theorem:main}. In Section~\ref{section:phasetransitions} we make
some general remarks on the expected phase transitions of our model, and in
Section~\ref{section:quartic} we study this in detail for the case of a quartic
potential.

\section{The equilibrium problem}
\label{section:equil}

\subsection{Existence of the minimizer}
\label{subsection:prooftheoremminimizer1}

In this section we prove the existence of the minimizer $(\mu_1,\mu_2)$ of the
equilibrium problem in Section~\ref{subsection:equilibriumproblem}. To this end
we follow \cite[Section 4]{DuK}.

\begin{proof}
The energy functional \eqref{eq:energyfunctional} can be written as
\begin{equation}\label{energyconvex}
E(\mu_1,\mu_2) =
    \frac{3}{4} I(\mu_1) + \frac{1}{4}I(\mu_1-2\mu_2)
        + \int \left( V(x) - a|x| \right) d\mu_1(x) \end{equation}
where
\[ I(\nu) = \iint \log \frac{1}{|x-y|} d\nu(x) d\nu(y) \]
denotes the logarithmic energy of a signed measure $\nu$. Occasionally we will
also write\[ I(\nu_1,\nu_2) = \iint \log \frac{1}{|x-y|} d\nu_1(x) d\nu_2(y)
\] to denote the mixed energy of a pair of measures $\nu_1$ and
$\nu_2$.

Since $I(\nu) \geq 0$ if $\nu$ is a signed measure with $\int d\nu = 0$, we
find from \eqref{energyconvex} that
\begin{align*}
    E(\mu_1,\mu_2) & \geq \frac{3}{4} I(\mu_1) + \int \left( V(x) - a |x| \right) d\mu_1(x) \\
        & \geq \min_{\mu_1} \left[ \frac{3}{4} I(\mu_1) + \int \left( V(x) - a |x| \right) d\mu_1(x) \right]
        > -\infty,
        \end{align*}
where the last inequality follows from standard logarithmic potential theory
with external fields, see e.g.\
\cite{SaffTotik}. Thus the energy functional is bounded from below.

If we fix $\mu_2$ on $i\mathbb R$ and we minimize \eqref{eq:energyfunctional}
with respect to $\mu_1$ only, then the problem for $\mu_1$ is to minimize
\[ I(\mu_1) + \int \left( V(x) - a|x| - U^{\mu_2}(x) \right) d\mu_1(x) \]
The extra term $U^{\mu_2}(x)$ comes from the interaction between $\mu_1$
and $\mu_2$. It is a term that attracts the $\mu_1$ mass towards
the origin. It can indeed be proved (as in \cite{DuK}) that
if the minimizer in external field $V(x) - a|x|$ is contained
in $[-X,X]$, then the minimizer in external field $V(x) - a|x| - U^{\mu_2}(x)$
is also contained in $[-X,X]$ (and so  $X$ is independent of $\mu_2$).

If we fix $\mu_1$ on $[-X,X]$ then the problem for $\mu_2$ is
to minimize
\[ I(\mu_2) - \int U^{\mu_1}(x) d\mu_2(x) \]
among all $\mu_2 \leq \sigma$ with total mass $1/2$. As in \cite{DuK}, equality
in the constraint is attained precisely on an interval of the form $[-ic,ic]$
for certain $c\geq 0$. We will show further that the minimizer $\mu_2$
satisfying this constraint is given explicitly by
\eqref{density:mu2a}--\eqref{density:mu2d}. From these explicit formulas it
follows immediately that for a measure $\mu_1$ on $[-X,X]$, the corresponding
minimizer $\mu_2$ satisfies
\[ \frac{d\mu_2(z)}{|dz|} \leq \frac{K}{|z|^2}, \qquad z \in i \mathbb R, \]
with a constant $K$ that only depends on $X$.

Let $(\mu_{1,n}, \mu_{2,n})_{n=1}^{\infty}$ be a sequence of vectors of
measures satisfying $\supp(\mu_{1,n}) \subset \mathbb R$, $\supp(\mu_{2,n})
\subset i\mathbb R$, $\int d\mu_{1,n} = 1$, $\int d\mu_{2,n} = 1/2$, and
$\mu_{2,n} \leq \sigma$, so that
\[ E(\mu_{1,n}, \mu_{2,n}) \leq \frac{1}{n} + \inf E(\mu_1,\mu_2). \]
As shown above, we may assume in addition that
\[ \supp(\mu_{1,n}) \subset [-X,X] \]
\[ \frac{d\mu_{2,n}(z)}{|dz|} \leq \frac{K}{|z|^2}, \qquad z \in i \mathbb R \]
with $X$ and $K$ independent of $n$.

Then it follows as in \cite{DuK} that the sequences
$(\mu_{1,n})_{n=1}^{\infty}$ and $(\mu_{2,n})_{n=1}^{\infty}$ are tight. There
is a convergent subsequence of $(\mu_{1,n}, \mu_{2,n})_{n=1}^{\infty}$ and the
limit is the vector of minimizing measures, see also \cite{KDragnev1}.

Summarizing, we have now proved the existence of the solution $(\mu_1,\mu_2)$
to the equilibrium problem. The uniqueness of the solution follows in a
standard way from the convexity of the energy functional, see e.g.\
\eqref{energyconvex} and \cite{DuK}.
\end{proof}

\subsection{Proof of Theorem~\ref{theorem:minimizer}}
\label{subsection:prooftheoremminimizer2}

In this section we prove Theorem~\ref{theorem:minimizer}.

\begin{proof}
The proof of Theorem~\ref{theorem:minimizer}(a) follows as in \cite{DKM}, while
Part~(c) is evident from the symmetry of the problem.

It remains to prove \eqref{condition:CaseI}--\eqref{density:mu2d} in
Theorem~\ref{theorem:minimizer}(b). For $k=1,2$, define the Cauchy transforms
\begin{equation} \label{def:Fj}
    F_k(z) = \int \frac{1}{z-s} \, d\mu_k(s), \qquad z \in \mathbb C \setminus
    \supp(\mu_k).
\end{equation}
By differentiating the variational condition \eqref{eq:varcondition3} we find
that
\begin{equation}\label{res2}
F_{2+}(z)+F_{2-}(z)- F_1(z)=0,\qquad z\in (-i\infty,-ic)\cup (ic,i\infty).
\end{equation}
Here we assume that the imaginary axis is oriented from bottom to top,
so that the $+$-side is on the left, and the $-$-side is on the right, as usual.
If $c>0$ then $\frac{d\mu_2}{|dz|} = \frac{d\sigma(z)}{|dz|} = \frac{a}{\pi}$
on $(-ic,ic)$, from which it follows that
\begin{equation}\label{res3}
    F_{2+}(z)-F_{2-}(z) = -2a, \qquad z\in (-ic,ic).
\end{equation}
We can solve equations (\ref{res2})--(\ref{res3}) for $F_2$. We consider the
two cases: $c=0$ and $c>0$.

{\it Case 1, $c=0$.} In this case we have equation (\ref{res2}) on $i\mathbb R$,
\begin{equation}\label{res2a}
F_{2+}(z)+F_{2-}(z)-F_1(z)=0,\qquad z\in (-i\infty,i\infty).
\end{equation}
Define the Cauchy transforms of the restrictions of the measure $\mu_1$ to the positive
and negative half-axes,
\begin{equation*}\label{res4}
\begin{aligned}
    F_1^+(z) & =\int_0^\infty \frac{d\mu_1(s)}{z-s},
    \qquad z\in\mathbb C\setminus \left(\mathbb{R}_+\cap\supp(\mu_1) \right),\\
    F_1^-(z) & =\int_{-\infty}^0 \frac{d\mu_1(s)}{z-s},\qquad z\in\mathbb C\setminus
    \left(\mathbb{R}_-\cap\supp(\mu_1) \right).
\end{aligned}
\end{equation*}
Then
\begin{equation*}\label{res5}
F_1(z)=F_1^+(z)+F_1^-(z),
\end{equation*}
and, due to the uniqueness of the solution of the scalar
Riemann-Hilbert problem  \eqref{res2a}, we have
\begin{equation*}\label{res6}
F_2(z)=
\begin{cases}
    F_1^+(z), & \quad \Re z\leq 0,\\
    F_1^-(z), & \quad \Re z\geq 0.
\end{cases}
\end{equation*}
Therefore, the Sokhotski-Plemelj formula implies that for $z \in i \mathbb R$,
\begin{equation*}\label{res6a}
\begin{aligned}
    \frac{d\mu_2(z)}{|dz|}&=-\frac{1}{2\pi } \left[F_{2+}(z)-F_{2-}(z)\right]=
    -\frac{1}{2\pi } \left[F_1^+(z)-F_1^-(z)\right]\\
    &=-\frac{1}{2\pi } \left[\int_0^\infty \frac{d\mu_1(s)}{z-s}-
    \int_0^\infty \frac{d\mu_1(s)}{z+s}\right] = \frac{1}{\pi }\int_0^\infty
    \frac{s\,d\mu_1(s)}{|z|^2 +s^2},
\end{aligned}
\end{equation*}
which is equivalent to \eqref{density:mu2a}. The case $c=0$ is valid
if and only if the density \eqref{density:mu2a} is bounded by $a/\pi$. Since
\eqref{density:mu2a} assumes its maximum for $|z| = 0$, this happens
if and only if \eqref{condition:CaseI} holds.

{\it Case 2, $c>0$.} Consider the function
\begin{equation*}\label{res7}
\widetilde F_2(z)=\frac{F_2(z)}{\sqrt{R(z)}},\qquad R(z)=z^2+c^2,
\end{equation*}
where $\sqrt{R(z)}=\sqrt{z^2+c^2}$ is defined with
a cut $(-i \infty,-ic]\cup[ic,i \infty)$, and $\sqrt{R(0)}=c$. From equation (\ref{res2})
we obtain that
\begin{equation}\label{res8}
    \widetilde F_{2+}(z)-\widetilde F_{2-}(z)=\frac{ F_1(z)}{\sqrt{R(z)}_+},\qquad
    z\in (-i\infty,-ic]\cup [ic,i\infty),
\end{equation}
where $+$ again denotes the limiting value from the left half plane.
Observe that for $y>c$,
\begin{equation*}\label{res8b}
\sqrt{R(iy)}_+=-i\sqrt{y^2-c^2}.
\end{equation*}
In addition, from equation (\ref{res3}) we obtain that
\begin{equation}\label{res9}
\widetilde F_{2+}(z)-\widetilde F_{2-}(z)=-\frac{2a}{\sqrt{R(z)}},\qquad
 z\in (-ic,ic).
\end{equation}
By \eqref{res8} and \eqref{res9}  the Sokhotski-Plemelj formula implies that
\begin{equation}\label{res10}
    \widetilde F_2(z)=-\frac{1}{2\pi i}\left(\int_{-i\infty}^{-ic}+\int_{ic}^{i\infty}\right)
    \frac{ F_1(s)\,ds}{(z-s)\sqrt{R(s)}_+}+\frac{1}{2\pi i}\int_{-ic}^{ic} \frac{ 2a\,ds}{(z-s)\sqrt{R(s)}}.
\end{equation}

\begin{center}
\vspace{-4mm}
\begin{figure}[t]
\begin{center}
\scalebox{0.50}{\includegraphics{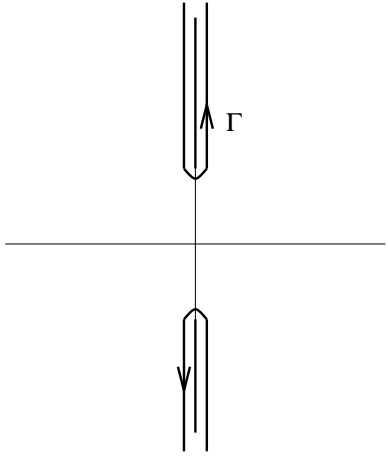}}
\end{center}
  \caption{The contour $\Gamma$.}
  \label{fig:contourGamma}
 \end{figure}
\end{center}

The first term in the right-hand side of \eqref{res10} can be written as
\begin{equation}\label{res11}
-\frac{1}{2\pi i}\left(\int_{-i\infty}^{-ic}+\int_{ic}^{i\infty}\right) \frac{
F_1(s)\,ds}{(z-s)\sqrt{R(s)}_+}=\frac{1}{4\pi i}\oint_{\Gamma} \frac{
F_1(s)\,ds}{(z-s)\sqrt{R(s)}},
\end{equation}
where the contour $\Gamma$ is depicted in Figure~\ref{fig:contourGamma}.
From \eqref{def:Fj} and Fubini's theorem,
\begin{multline}
    \frac{1}{4\pi i}\oint_{\Gamma} \frac{F_1(s)\,ds}{(z-s)\sqrt{R(s)}} =
    \frac{1}{4\pi i}\int_{-\infty}^\infty \oint_{\Gamma} \frac{ ds}{(z-s)(s-\zeta)\sqrt{R(s)}}\,d\mu_1(\zeta)
    \\ = \frac{1}{4\pi i}\int_{-\infty}^\infty\frac{1}{z-\zeta}
    \left(\oint_{\Gamma}\frac{ ds}{(z-s)\sqrt{R(s)}} +
\oint_{\Gamma}\frac{ ds}{(s-\zeta)\sqrt{R(s)}}\right)d\mu_1(\zeta). \label{res13}
\end{multline}
By contour deformation and Cauchy's theorem we have that
\begin{equation*}\label{res14}
\begin{aligned}
\frac{1}{4\pi i} \oint_{\Gamma}\frac{
ds}{(z-s)\sqrt{R(s)}}=\frac{1}{2\sqrt{R(z)}}, \quad z\in \mathbb C\setminus
    ((-i \infty,-ic]\cup[ic,i\infty)),
\end{aligned}
\end{equation*}
hence \eqref{res13} reduces to
\begin{multline}
    \frac{1}{4\pi i}\oint_{\Gamma} \frac{ F_1(s)\,ds}{(z-s)\sqrt{R(s)}}
    =\int_{-\infty}^\infty\frac{1}{2(z-\zeta)}
    \left(\frac{1}{\sqrt{R(z)}}-\frac{1}{\sqrt{R(\zeta)}}\right)\,d\mu_1(\zeta)
    \\ \label{res16}
    = \frac{1}{\sqrt{R(z)}}\int_{-\infty}^\infty
\frac{d\mu_1(\zeta)}{2(z-\zeta)}\, -\int_{-\infty}^\infty
\frac{d\mu_1(\zeta)}{2(z-\zeta)\sqrt{R(\zeta)}}.
\end{multline}

For the second  term in the right-hand side of \eqref{res10} we have
\begin{equation}\label{res17}
    \frac{1}{2\pi i}\int_{-ic}^{ic} \frac{ 2a\,ds}{(z-s)\sqrt{R(s)}}
    =\frac{a\,\sign \Re z}{ \sqrt{R(z)}}.
\end{equation}
By inserting \eqref{res16} and \eqref{res17} in \eqref{res10}, we obtain that
\begin{equation*}\label{res18}
\tilde F_2(z)=\frac{1}{\sqrt{R(z)}}\int_{-\infty}^\infty
\frac{d\mu_1(s)}{2(z-s)}\, -\int_{-\infty}^\infty
\frac{d\mu_1(s)}{2(z-s)\sqrt{R(s)}}+\frac{a\,\sign \Re z}{ \sqrt{R(z)}},
\end{equation*}
hence
\begin{equation}\label{res19}
F_2(z)=\frac{F_1(z)}{2} -\sqrt{z^2+c^2}\int_{-\infty}^\infty
\frac{d\mu_1(s)}{2(z-s)\sqrt{s^2+c^2}}+a\,\sign \Re z.
\end{equation}
Note that by taking $z\to +\infty$ in \eqref{res19}, we find the relation
\eqref{condition:CaseIIorIII} between $c$ and $a$.
Now the density of $\mu_2$ is equal to
\begin{equation*}\label{res22}
\begin{aligned}
\frac{d\mu_2(z)}{|dz|}&=-\frac{1}{2\pi }\left[ F_{2+}(z)-F_{2-}(z)\right] \\
&=\frac{a}{\pi}-\frac{i\sqrt{|z|^2-c^2}}{\pi}\int_{-\infty}^\infty
\frac{d\mu_1(s)}
{2(z-s)\sqrt{s^2+c^2}}\\
&=\frac{a}{\pi}-\frac{i\sqrt{|z|^2-c^2}}{\pi}\int_0^\infty
\left(\frac{1}{z-s}+\frac{1}{z+s}\right) \frac{d\mu_1(s)}
{2\sqrt{s^2+c^2}}\\
&=\frac{a}{\pi}-\frac{1}{\pi }\int_0^\infty \frac{|z| \sqrt{|z|^2-c^2}}{|z|^2+s^2}
\frac{d\mu_1(s)} {\sqrt{s^2+c^2}},
\end{aligned}
\end{equation*}
which is equivalent to \eqref{density:mu2c}. Then \eqref{density:mu2d} follows
from this and \eqref{condition:CaseIIorIII}.
\end{proof}

\subsection{Structure of the equilibrium measures in the regular case}
\label{subsection:threecases}

Theorem~\ref{theorem:minimizer} implies that in the regular case, the structure
of the equilibrium measures near the origin is described by one of the
following three cases. This distinction will be important at several places of
our RH steepest descent analysis.

\begin{itemize}
\item[] \textbf{Case I}: $c=0$ and $N$ is even.
\item[] \textbf{Case II}: $c>0$ and $N$ is even.
\item[] \textbf{Case III}: $c>0$ and $N$ is odd.
\end{itemize}
Here we recall the definitions of $c$ and $N$ in \eqref{def:N}--\eqref{def:c}.
Thus Case~I could be formulated equivalently as $\supp(\sigma-\mu_2) = i
\mathbb R$ and $0 \not\in \supp(\mu_1)$. Similar formulations can be given for
Case~II and Case~III. Note that the situation where $c=0$ and $N$ is odd is not in
the above list since it would imply that $\mu_1$ has singular behavior at the
origin.

For the quadratic potential $V(x)=x^2/2$, it turns out that we are in Case~I
(with $N=2$) for large values of $a$ and in Case~III (with $N=1$) for small
values of $a$. The Case~II does not occur.

In general, one expects Case~I to happen for large values of $a$. Note that the
upper constraint $\sigma$ is not active in this case and therefore it could be
removed from the equilibrium problem. A consequence of this is that $2\mu_2$ is
equal to the balayage of $\mu_1$ onto the imaginary axis, see also \eqref{density:mu2a}.
This means that $2 U^{\mu_2}(z) = U^{\mu_1}(z)$ for $z \in i \mathbb R$.

Let $\mu_1^L$ and $\mu_1^R$ denote the restrictions of $\mu_1$ to the negative
and positive real axis, respectively. By symmetry, we then have in Case~I that
$\mu_2$ is the balayage of either $\mu_1^L$ or $\mu_1^R$ onto the imaginary
axis, and moreover
\begin{align}
    U^{\mu_2}(z) & = U^{\mu_1^L}(z), \qquad \text{for } \Re z \geq 0, \\
    U^{\mu_2}(z) & = U^{\mu_1^R}(z), \qquad \text{for } \Re z \leq 0.
    \end{align}
Note that the equality is valid not only on the imaginary axis, but also in a
full half-plane. This follows from an easy application of the minimum and
maximum principles for harmonic functions \cite[Chapter 0]{SaffTotik}.

Then the following string of equations is easy to verify:
\begin{align} \nonumber
    I(\mu_2) & = \int U^{\mu_2} d\mu_2 = \int U^{\mu_1^L} d\mu_2 = \int U^{\mu_2} d\mu_1^L \\
        & = \int U^{\mu_1^R} d\mu_1^L
        \label{towardsAngelesco1} = I(\mu_1^L, \mu_1^R),
        \end{align}
and in a similar way
\begin{align}
    I(\mu_1,\mu_2) & = \int U^{\mu_2} d\mu_1^L + \int U^{\mu_2} d\mu_1^R
       \label{towardsAngelesco2}  = 2 I(\mu_1^L, \mu_1^R).
        \end{align}
Since also
\begin{align}\label{towardsAngelesco3}
    I(\mu_1) & = I(\mu_1^L + \mu_1^R) =
        I(\mu_1^L) + I(\mu_1^R) + 2 I(\mu_1^L, \mu_1^R),
        \end{align}
it then follows from \eqref{towardsAngelesco1}--\eqref{towardsAngelesco3} that
the energy functional \eqref{eq:energyfunctional} can be rewritten in Case I as
\begin{multline*} E(\mu_1, \mu_2) =
        I(\mu_1^L) + I(\mu_1^R) + I(\mu_1^L, \mu_1^R) \\
    + \int \left(V(x) + a x\right) d\mu_1^L(x)
    + \int \left(V(x) - a x\right) d\mu_1^R(x).
\end{multline*}
Therefore the  equilibrium problem is equivalent to the
following equilibrium problem of Angelesco type for $\mu_1^L$ and $\mu_1^R$:
Minimize
\begin{multline}
    E(\mu_1^L,\mu_1^R) = \iint \log \frac{1}{|x-y|} d\mu_1^L(x) d\mu_1^L(y)
    + \iint \log \frac{1}{|x-y|} d\mu_1^R(x) d\mu_1^R(y) \\
    + \iint \log \frac{1}{|x-y|} d\mu_1^L(x) d\mu_1^R(y) \\
    + \int \left(V(x) + a x\right) d\mu_1^L(x)
    + \int \left(V(x) - a x\right) d\mu_1^R(x),
    \end{multline}
with respect to all pairs of measures $(\mu_1^L,\mu_1^R)$ satisfying
\begin{itemize}
\item $\mu_1^L$ is a measure on $\mathbb R_{-}$ with total mass $1/2$
\item $\mu_1^R$ is a measure on $\mathbb R_{+}$ with total mass $1/2$.
\end{itemize}
For the quadratic potential $V(x) = \frac{1}{2}x^2$,
this equilibrium problem was described in
\cite{BK2}.

\begin{figure}[t]
\begin{center}
   \setlength{\unitlength}{1truemm}
   \begin{picture}(100,50)(-5,2)

       \put(7,29){\line(1,0){55}}
       \put(19,41){\line(1,0){55}}
       \put(7,29){\line(1,1){12}}
       \put(62,29){\line(1,1){12}}
       \thicklines
       \put(20,35){\line(1,0){8}}
       \put(32,35){\line(1,0){7}}
       \put(44,35){\line(1,0){7}}
       \put(55,35){\line(1,0){8}}
       \thinlines
       \put(75,35){$\mathcal R_1$}

       \put(7,14){\line(1,0){55}}
       \put(19,26){\line(1,0){55}}
       \put(7,14){\line(1,1){12}}
       \put(62,14){\line(1,1){12}}
       \thicklines
       \put(44,20){\line(1,0){7}}
       \put(55,20){\line(1,0){8}}
       \thinlines
       \put(75,20){$\mathcal R_2$}

       \put(7,-1){\line(1,0){55}}
       \put(19,11){\line(1,0){55}}
       \put(7,-1){\line(1,1){12}}
       \put(62,-1){\line(1,1){12}}
       \thicklines
       \put(20,5){\line(1,0){8}}
       \put(32,5){\line(1,0){7}}
       \thinlines
       \put(75,5){$\mathcal R_3$}

   \end{picture}
   \vspace{1mm}
   \caption{Riemann surface for Case I (with $N=4$).}
   \label{fig:Riemannsurface1}
\end{center}
\end{figure}
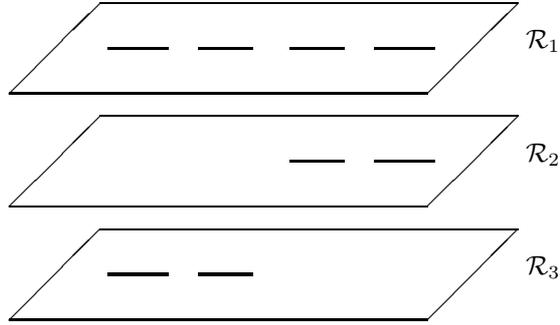

\section{Riemann surface}
\label{section:riemannsurface}

From the minimizer $(\mu_1,\mu_2)$ of the vector equilibrium problem we
construct a three sheeted Riemann surface $\mathcal R$, whose three sheets are
given as follows.
\begin{align} \label{eq:sheetR1}
    \mathcal R_1 & =  \overline{\mathbb C} \setminus \bigcup_{j=1}^N [a_j, b_j], \\
    \label{eq:sheetR2}
    \mathcal R_2 & = \overline{\mathbb C} \setminus
    \left( \left(\mathbb{R}_+\cap\bigcup_{j=1}^N [a_j, b_j] \right) \cup \overline{(-ic,ic)} \right), \\
    \label{eq:sheetR3}
    \mathcal R_3 & = \overline{\mathbb C} \setminus
    \left( \left(\mathbb{R}_-\cap\bigcup_{j=1}^N [a_j, b_j] \right) \cup \overline{(-ic,ic)}\right).
    \end{align}
    Here $\overline{\mathbb C} = \mathbb C\cup\{\infty\}$ denotes the Riemann
    sphere.

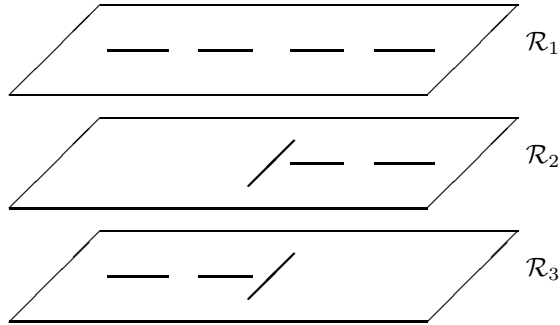
\begin{figure}[t]
\begin{center}
   \setlength{\unitlength}{1truemm}
   \begin{picture}(100,50)(-5,2)

       \put(7,29){\line(1,0){55}}
       \put(19,41){\line(1,0){55}}
       \put(7,29){\line(1,1){12}}
       \put(62,29){\line(1,1){12}}
       \thicklines
       \put(20,35){\line(1,0){8}}
       \put(32,35){\line(1,0){7}}
       \put(44,35){\line(1,0){7}}
       \put(55,35){\line(1,0){8}}
       \thinlines
       \put(75,35){$\mathcal R_1$}

       \put(7,14){\line(1,0){55}}
       \put(19,26){\line(1,0){55}}
       \put(7,14){\line(1,1){12}}
       \put(62,14){\line(1,1){12}}
       \thicklines
       \put(44,20){\line(1,0){7}}
       \put(55,20){\line(1,0){8}}
       \put(38.5,17){\line(1,1){6}}
       \thinlines
       \put(75,20){$\mathcal R_2$}

       \put(7,-1){\line(1,0){55}}
       \put(19,11){\line(1,0){55}}
       \put(7,-1){\line(1,1){12}}
       \put(62,-1){\line(1,1){12}}
       \thicklines
       \put(20,5){\line(1,0){8}}
       \put(32,5){\line(1,0){7}}
       \put(38.5,2){\line(1,1){6}}
       \thinlines
       \put(75,5){$\mathcal R_3$}

   \end{picture}
   \vspace{1mm}
   \caption{Riemann surface for Case II (with $N=4$).}
   \label{fig:Riemannsurface2}
\end{center}
\end{figure}

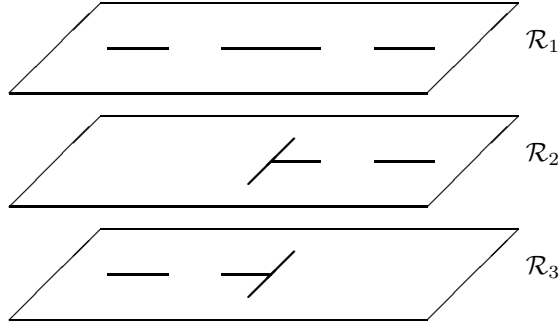
\begin{figure}[t]
\begin{center}
   \setlength{\unitlength}{1truemm}
   \begin{picture}(100,50)(-5,2)

       \put(7,29){\line(1,0){55}}
       \put(19,41){\line(1,0){55}}
       \put(7,29){\line(1,1){12}}
       \put(62,29){\line(1,1){12}}
       \thicklines
       \put(20,35){\line(1,0){8}}
       \put(35,35){\line(1,0){13}}
       \put(55,35){\line(1,0){8}}
       \thinlines
       \put(75,35){$\mathcal R_1$}

       \put(7,14){\line(1,0){55}}
       \put(19,26){\line(1,0){55}}
       \put(7,14){\line(1,1){12}}
       \put(62,14){\line(1,1){12}}
       \thicklines
       \put(41.5,20){\line(1,0){6.5}}
       \put(55,20){\line(1,0){8}}
       \put(38.5,17){\line(1,1){6}}
       \thinlines
       \put(75,20){$\mathcal R_2$}

       \put(7,-1){\line(1,0){55}}
       \put(19,11){\line(1,0){55}}
       \put(7,-1){\line(1,1){12}}
       \put(62,-1){\line(1,1){12}}
       \thicklines
       \put(20,5){\line(1,0){8}}
       \put(35,5){\line(1,0){6.5}}
       \put(38.5,2){\line(1,1){6}}
       \thinlines
       \put(75,5){$\mathcal R_3$}

   \end{picture}
   \vspace{1mm}
   \caption{Riemann surface for Case III (with $N=3$).}
   \label{fig:Riemannsurface3}
\end{center}
\end{figure}

The sheet $\mathcal R_1$ is connected with $\mathcal R_2$ via the intervals
$[a_j,b_j]$ on the positive real line, $\mathcal R_1$ is connected with
$\mathcal R_3$ via the intervals $[a_j,b_j]$ on the negative real line, and
(in Case~II and Case~III) $\mathcal R_2$ is connected to $\mathcal R_3$ via the interval $[-ic, ic]$ on
the imaginary axis. The connections are in the usual crosswise manner. The
Riemann surface is compact and has genus
\begin{equation} \label{eq:genusR}
    g = \left\{ \begin{array}{ll}
    N - 2, & \textrm{ in Case~I}, \\
    N - 1, & \textrm{ in Case~II and Case~III}. \end{array} \right.
    \end{equation}
Here the Cases I, II and III were defined in
Section~\ref{subsection:threecases}. An illustration of the Riemann surface for
each of these three cases is shown in
Figures~\ref{fig:Riemannsurface1}--\ref{fig:Riemannsurface3}.

Recall the functions $F_1$ and $F_2$ in \eqref{def:Fj}. These functions are
used to define a meromorphic function on the Riemann surface, compare with
\cite[Lemma 5.1]{DuK}:

\begin{proposition} \label{prop:xifunction}
For $k=1,2,3$, let $\xi_k(z)$ on the sheet $\mathcal R_k$ be defined by
\begin{equation}\label{def:xi}
\begin{aligned}
    \xi_1(z) & = V'(z) - F_1(z), \qquad\quad\ \ \! z \in \mathcal R_1, \\
    \xi_2(z) & = a + F_1(z) - F_2(z), \qquad\! z \in \mathcal R_2, \quad \Re z > 0, \\
    \xi_2(z) & = a + F_2(z), \qquad\qquad\quad\  z \in \mathcal R_2, \quad \Re z < 0, \\
    \xi_3(z) & = -a + F_2(z), \qquad\qquad\quad\!\! z \in \mathcal R_3, \quad \Re z > 0, \\
    \xi_3(z) & = -a + F_1(z) - F_2(z), \quad\ \! z \in \mathcal R_3, \quad \Re z < 0.
\end{aligned}
\end{equation}
Then these functions have an analytic continuation to a meromorphic function
(denoted by $\xi(z)$) on the Riemann surface whose only pole is at the point at
infinity on the first sheet.
\end{proposition}

\begin{proof} Let us first check that $\xi_k$ is analytic on the
sheet $\mathcal R_k$, $k=2,3$. This reduces to showing the equality
$$ F_{2,+}(z)+ F_{2,-}(z)-F_1(z)  = 0,\qquad z\in i\mathbb R\setminus
[-ic,ic],$$ which is a direct consequence of the variational condition
\eqref{eq:varcondition3}, see also \eqref{res2}.

Next, we must show that the $\xi$-functions are each others analytic
continuation when passing through one of the cuts of the Riemann surface. For
example, the fact that $\xi_{1,+}=\xi_{2,-}$ along the cut
$\mathbb{R}_+\cap\bigcup_{j=1}^N [a_j, b_j]$ reduces to showing the equality
$$ F_{1,+}(x)+F_{1,-}(x) -F_2(x) - V'(x) + a =0,\qquad x\in \mathbb R_+\cap\bigcup_{j=1}^N (a_j,
b_j),
$$
which is a direct consequence of the variational condition
\eqref{eq:varcondition1}. The other equalities are checked similarly, see also
\eqref{res3}.
\end{proof}

It follows from Proposition \ref{prop:xifunction} that
the function $\xi(z)$ is an algebraic function satisfying  an equation of the
third degree in $\xi$, known as the spectral curve:
\begin{equation} \label{eq:spectralcurve}
    \xi^3 + p_2(z) \xi^2 + p_1(z) \xi + p_0(z) = 0
    \end{equation}
    where $p_0$, $p_1$, $p_2$ are polynomials.
Here
\begin{equation} \label{eq:spectralcurvep2}
    p_2(z) = - \xi_1(z) - \xi_2(z) - \xi_3(z) = -V'(z)
    \end{equation}
is known, but the determination of the polynomials
\begin{align*}
    p_1(z) & = \xi_1(z) \xi_2(z) + \xi_1(z) \xi_3(z) + \xi_2(z) \xi_3(z) \\
    p_0(z) & = - \xi_1(z) \xi_2(z) \xi_3(z)
    \end{align*}
cannot be done in general. We can only certify that (we use that $V(z)$ is a
polynomial of degree $2d$ and $F_1(z) = 1/z + O(1/z^3)$, $F_2(z) = 1/(2z) +
o(1/z)$ as $z \to \infty$)
\begin{align*}
    p_1(z) & = V'(z) F_1(z) - a^2  + \mathcal O\left(z^{-2}\right) \\
        & = \frac{V'(z)}{z} - a^2  + \mathcal O\left(z^{2d-4}\right) \\
    p_0(z) & = a^2 V'(z) - \frac{V'(z)}{4z^2} + \mathcal O\left(z^{2d-3}\right)
\end{align*}
as $z \to \infty$.

\begin{example}
For $d=1$ we are in the quadratic case. If
\[ V(z) = \frac{1}{2} z^2 \]
then $p_1(z) = 1 - a^2$, $p_0(z) = a^2 z$ so that the spectral curve is
\begin{equation} \label{eq:Pastur}
    \xi^3 - z \xi^2 + (1-a^2) \xi + a^2 z = 0.
    \end{equation}
This is known as Pastur's equation \cite{Pastur}. It plays an important role in
\cite{ABK,BK2,BK3}.
\end{example}

\begin{example}
For $d=2$ we are in the quartic case. Let's take
\[ V(z) = \frac{1}{4} z^4 - \frac{t}{2} z^2. \]
Then $p_1(z) = z^2 + \mathcal O(1)$ and $p_0(z) = a^2 z^3 + \mathcal O(z)$ as
$z \to \infty$. By symmetry we have that $p_1$ is an even polynomial and $p_0$
is an odd polynomial, so that $p_1$ and $p_0$ are determined up to two
constants. Thus for some $\alpha, \beta \in \mathbb R$, we have
\[ p_1(z) = z^2 + \alpha, \qquad p_0(z) = a^2 z^3 + \beta z, \]
and the spectral curve is
\begin{equation} \label{eq:McLaughlin}
    \xi^3 - \left(z^3 - tz\right) \xi^2 + (z^2 + \alpha) \xi + a^2 z^3 + \beta z = 0.
    \end{equation}
This is McLaughlin's equation, named after K.T-R McLaughlin \cite{McL} who
derived it first for the case $t=0$, see also \cite{ALT}. We will analyze
this case in more detail in Section \ref{section:quartic} below.
\end{example}

\section{Proof of Theorem \ref{theorem:main}}
\label{section:steepestdescent}

Recall the RH problem for $Y(z)$ in \eqref{eq:Yjump}--\eqref{eq:Yatinfty}. In
Subsections~\ref{subsection:transfo1}--\ref{subsection:transfofinal} we will
perform a Deift-Zhou steepest descent analysis of this RH problem. This will
then lead to the proof of Theorem~\ref{theorem:main} in
Section~\ref{subsection:proofmaintheorem}.

\subsection{First transformation $Y \mapsto X$}
\label{subsection:transfo1}

In the first transformation we open up an unbounded lens around $\supp(\sigma-\mu_2)$
which is bounded by a contour $\Gamma = \Gamma^+ \cup \Gamma^-$.
We choose the lens so that it is symmetric
under reflection with respect to both the real and the imaginary axis.
The construction of the lens depends on whether we are in Case I or in
one of the other cases (Case II or Case III).

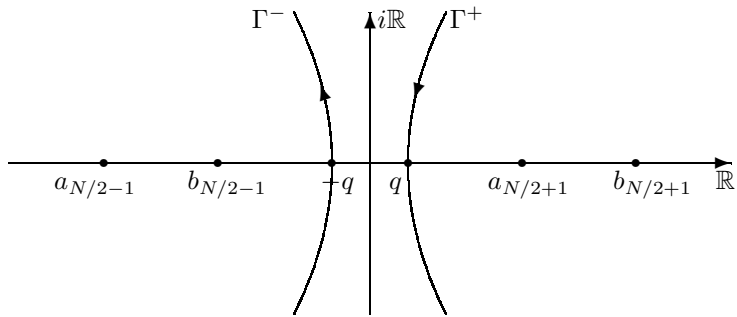
\begin{figure}[t]
\begin{center}
   \setlength{\unitlength}{1truemm}
   \begin{picture}(100,70)(-5,2)

       \put(0,40){\line(1,0){95}}
       \put(47.5,20){\line(0,1){40}}
       \put(93,36.6){$\mathbb R $}
       \put(48.5,58){$i\mathbb R $}
       \put(95,40){\thicklines\vector(1,0){.0001}}
       \put(47.5,60){\thicklines\vector(0,1){.0001}}
       \put(53.3,48.5){\thicklines\vector(-1,-3){.0001}}
       \put(41,50.5){\thicklines\vector(-1,3){.0001}}
       \put(42.5,40){\thicklines\circle*{1}}
       \put(52.5,40){\thicklines\circle*{1}}
       \put(67.5,40){\thicklines\circle*{1}}
       \put(82.5,40){\thicklines\circle*{1}}
       \put(50,36.6){$q$}
       \put(41,36.6){$-q$}
       \put(63,36.6){$a_{N/2+1}$}
       \put(79.5,36.6){$b_{N/2+1}$}
       \put(12.5,40){\thicklines\circle*{1}}
       \put(27.5,40){\thicklines\circle*{1}}
       \put(6,36.6){$a_{N/2-1}$}
       \put(23.5,36.6){$b_{N/2-1}$}

       \qbezier(37.5,60)(47.5,40)(37.5,20)
       \qbezier(57.5,60)(47.5,40)(57.5,20)
       \put(58,58){$\Gamma^{+}$}

       \put(32,58){$\Gamma^{-}$}
   \end{picture}
 \vspace{-20mm}
   \caption{Lens around $\textrm{Supp}(\sigma-\mu_2) = i\mathbb R$ in Case I.}
   \label{fig:globallens1}
\end{center}
\end{figure}

First assume that we are in Case I so that $\supp(\sigma - \mu_2) = i \mathbb
R$. Then we take $\Gamma$ as in Figure~\ref{fig:globallens1}. That is, we take
$q \in (0, a_{N/2}+1)$ and let $\Gamma^+$ be an unbounded Jordan curve in the
right half-plane, symmetric in the real axis and crossing the real axis in $q$.
Also $\Gamma^+$ is asymptotic to the half-ray $\arg z = \pm \theta$ for some
$\theta \in (0, \pi/2)$. Then $\Gamma^-$ is the reflection of $\Gamma^+$ in the
imaginary axis. We orient $\Gamma$ as shown in Figure~\ref{fig:globallens1}.

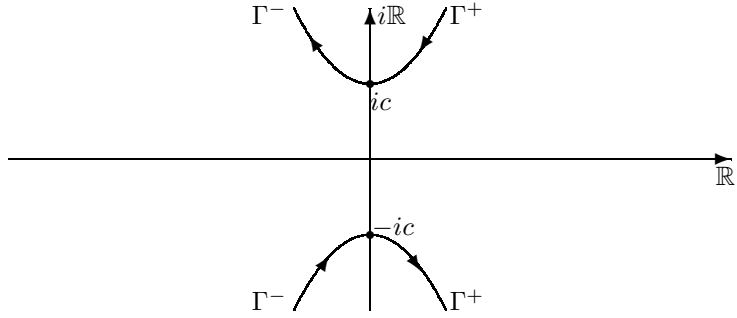
\begin{figure}[t]
\begin{center}
   \setlength{\unitlength}{1truemm}
   \begin{picture}(100,70)(-5,2)

       \put(0,40){\line(1,0){95}}
       \put(47.5,20){\line(0,1){40}}
       \put(93,36.6){$\mathbb R $}
       \put(48.5,58){$i\mathbb R $}
       \put(95,40){\thicklines\vector(1,0){.0001}}
       \put(47.5,60){\thicklines\vector(0,1){.0001}}
       \put(54.0,54.3){\thicklines\vector(-2,-3){.0001}}
       \put(54.2,25.2){\thicklines\vector(2,-3){.0001}}
       \put(39.5,56.4){\thicklines\vector(-2,3){.0001}}
       \put(42.15,27.2){\thicklines\vector(2,3){.0001}}
       \put(47.5,50){\thicklines\circle*{1}}
       \put(47.5,46.6){$ic$}
       \put(47.5,30){\thicklines\circle*{1}}
       \put(47.8,30){$-ic$}

       \qbezier(37.5,60)(47.5,40)(57.5,60)
       \qbezier(37.5,20)(47.5,40)(57.5,20)
       \put(58,58){$\Gamma^{+}$}
       \put(32,58){$\Gamma^{-}$}

       \put(58,20){$\Gamma^{+}$}
       \put(32,20){$\Gamma^{-}$}
   \end{picture}
 \vspace{-20mm}
   \caption{Lens around $\textrm{Supp}(\sigma-\mu_2)$ in Cases II and III.}
   \label{fig:globallens23}
\end{center}
\end{figure}

Next, consider the Cases II and III. Then we take $\Gamma = \Gamma^+ \cup \Gamma^-$
as in Figure~\ref{fig:globallens23}. The part of $\Gamma^+$ in the upper half
plane is  a Jordan curve going from $\infty$ at an angle $\theta \in (0, \pi/2)$
to the point $ic$. The other part of $\Gamma^+$ is its reflection in the real
axis, and $\Gamma^-$ is obtained from $\Gamma^+$ by reflection in the imaginary
axis.  We orient $\Gamma$ as shown
in Figure~\ref{fig:globallens23}.

The precise way to choose the contour $\Gamma$ will be described further on.

The contour $\Gamma$ divides the complex plane into an inner and an outer part. By
definition, we say that $\supp(\sigma-\mu_2)$ is inside the lens, while
$\supp(\mu_1)$ is outside the lens.
Note that our definitions are such that the outside of the lens is
always on the left when traversing $\Gamma$ according to its orientation.

We define a new $3\times 3$ matrix valued function $X$ by
\begin{equation}\label{def:X}
    X(z) = \left\{\begin{array}{ll} Y(z) \begin{pmatrix} 1 & 0 & 0 \\ 0 & 1 & -e^{-2naz} \\ 0 & 0 & 1
    \end{pmatrix}, &
    \qquad \textrm{for } \Re z > 0 \textrm{ outside the lens}, \\
    Y(z) \begin{pmatrix} 1 & 0 & 0 \\ 0 & 1 & 0 \\ 0 & -e^{2naz} & 1
    \end{pmatrix},
    & \qquad \textrm{for } \Re z < 0 \textrm{ outside the lens},\\
    Y(z),& \qquad \textrm{for $z$ inside the lens}.
    \end{array}\right.
    \end{equation}

Recall that $q$ is used to denote the intersection of $\Gamma^+$ with
the real axis (in Case I). In Cases II and III we put $q=0$.

Then $X$ satisfies the following RH problem.

\begin{itemize}
\item[(1)] $X$ is analytic in $\mathbb C \setminus (\mathbb R \cup [-ic,ic] \cup \Gamma)$.
\item[(2)] $X$ satisfies the jump properties
\begin{align}
    X_+(x) & = X_-(x) \begin{pmatrix} 1 & e^{-n(V(x)-ax)} & 0 \\ 0 & 1 & 0 \\ 0 & 0 & 1 \end{pmatrix}
    \qquad \textrm{for } x > q, \\
    X_+(x) & = X_-(x) \begin{pmatrix} 1 & 0 & e^{-n(V(x)+ax)} \\ 0 & 1 & 0 \\ 0 & 0 & 1 \end{pmatrix}
    \qquad \textrm{for } x < -q,\\
    X_+(x) & = X_-(x) \begin{pmatrix} 1 & e^{-n(V(x)-ax)} & e^{-n(V(x)+ax)} \\ 0 & 1 & 0 \\ 0 & 0 & 1 \end{pmatrix}
    \quad \textrm{for } -q < x < q \textrm{ (in Case I)},
\end{align}
on the real line,
\begin{align}
    X_+(z) & = X_-(z) \begin{pmatrix} 1 & 0 & 0 \\ 0 & 0 & e^{-2naz} \\ 0 & -e^{2naz} & 1 \end{pmatrix}
    \quad \textrm{for } z \in (-ic,ic) \textrm{ (in Case II, III)},
\end{align}
on the interval $(-ic,ic)$ oriented upwards, and
\begin{align}
    X_+(z) & = X_-(z) \begin{pmatrix} 1 & 0 & 0 \\ 0 & 1 & -e^{-2naz} \\ 0 & 0 & 1 \end{pmatrix}
    \qquad \textrm{for } z\in\Gamma^+, \\
    X_+(z) & = X_-(z) \begin{pmatrix} 1 & 0 & 0 \\ 0 & 1 & 0 \\ 0 & -e^{2naz} & 1 \end{pmatrix}
    \qquad \textrm{for } z\in\Gamma^-,
\end{align}
on the contour $\Gamma$.

\item[(c)] As $z\to\infty$ we have
\begin{align}
    X(z) = \left( I + O(1/z)\right)
    \begin{pmatrix} z^n & 0 & 0 \\ 0 & z^{-n/2} & 0 \\ 0 & 0 & z^{-n/2}
    \end{pmatrix}. \end{align}
\end{itemize}

\subsection{Second transformation $X \mapsto T$}
\label{subsection:transfo2}

In the second transformation $X \mapsto T$ we use the minimizers $\mu_1$ and
$\mu_2$ of the vector equilibrium problem, and the associated $g$-functions
\begin{equation}\label{def:g} g_k(z) = \int \log(z-s) d \mu_k(s). \end{equation}

For $k=1$ we choose the branch of the logarithm in \eqref{def:g} in the
standard way, with a branch cut $(-\infty,s]$ along the negative real axis.
Thus $g_1(z)$ is defined and analytic in $\mathbb C \setminus (-\infty, b_N]$
and
\[ g_1(z) = \log z + O(1/z), \qquad \textrm{as } z \to \infty. \]

For $k=2$ the logarithm in \eqref{def:g} is taken with branch cut
 $(-i\infty, s]$ along the imaginary axis. Thus
 $\log(z-s) = \log|z-s| + i \arg(z-s)$ with $-\pi/2 < \arg(z-s) < 3\pi/2$.
Then $g_2(z)$ is defined and analytic in $\mathbb C \setminus i \mathbb R$, and
\[ g_2(z) = \frac{1}{2} \log z + O(1), \qquad \textrm{as } z \to \infty,  \]
with the above branch of the logarithm.

The behavior of the real and imaginary parts of the $g$-functions is described
in the following lemma.

\begin{lemma}\label{lemma:gfunctions}
For $x\in\mathbb R\setminus\{0\}$ we have
\begin{equation}\label{eq:g1}
\begin{aligned}
g_{1,+}(x)+g_{1,-}(x) &= -2U^{\mu_1}(x) \\
g_{1,+}(x)-g_{1,-}(x) &= 2\pi i \mu_1([x,+\infty)) \\
g_{2}(x) &= -U^{\mu_2}(x)+\left\{\begin{array}{ll} 0,\qquad\ x>0\\
\pi i/2,\quad x<0.\end{array}\right.
\end{aligned}
\end{equation}
For $z\in i\mathbb R\setminus\{0\}$ we have
\begin{equation}\label{eq:g2}
\begin{aligned} g_{2,+}(z)+g_{2,-}(z) &= -2U^{\mu_2}(z) +\pi i/2\\
g_{2,+}(z)-g_{2,-}(z) &= \left\{\begin{array}{ll} 2\pi i\mu_2([z,i\infty )),\qquad z\in i\mathbb R \\
\pi i/2 - 2az,\qquad\quad\ z\in [-ic,ic]
\end{array}\right. \\
g_1(z) &= -U^{\mu_1}(z)+\left\{\begin{array}{ll} \pi i/2, \qquad\ \ z\in i\mathbb R_+\\
-\pi i/2, \qquad z\in i\mathbb R_-. \end{array}\right.
\end{aligned}
\end{equation}
\end{lemma}

\begin{proof}
The equalities follow immediately from the definitions, where we have to be
careful with the choice of branches of the
logarithm as discussed above.

Only the equality in \eqref{eq:g2} for $z \in [-ic,ic]$ needs some extra comment.
We have that $\mu_2 = \sigma$ on $[-ic,ic]$ and so if $z \in [0,ic]$,
\[ \mu_2([0,z]) = \sigma([0,z]) = \frac{a}{\pi} |z| = - \frac{a}{\pi} i z, \]
and therefore
\begin{equation} \label{eq:g2proof}
    \mu_2([z,i\infty)) = \mu_2([0,i\infty)) - \mu_2([0,z])
    = \frac{1}{4} + \frac{a}{\pi} i z,
    \end{equation}
since $\mu_2$ is symmetric in the origin and $\mu_2(i \mathbb R) = 1/2$.
The equality \eqref{eq:g2proof} is proved for $z \in [-ic, 0]$ in a similar way,
and thus
\[ g_{2,+}(z) - g_{2,-}(z) = 2\pi i \mu_2([z,i\infty)) =
    \pi i/2 - 2a z, \qquad z \in [-ic,ic]. \]
\end{proof}

Now we introduce the $\lam$-functions, which are defined as the following
anti-derivatives of the $\xi$-functions \eqref{def:xi}.
Recall that $\ell$ is the constant in the variational conditions
\eqref{eq:varcondition1}--\eqref{eq:varcondition2}.

\begin{equation}\label{def:lambda}
\begin{aligned}
    \lam_1(z) & = V(z) + \ell - g_1(z), \qquad\quad \! z \in \mathcal R_1, \\
    \lam_2(z) & = az + g_1(z) - g_2(z), \qquad\! z \in \mathcal R_2, \quad \Re z > 0, \\
    \lam_2(z) & = az + g_2(z), \qquad\qquad\quad\  z \in \mathcal R_2, \quad \Re z < 0, \\
    \lam_3(z) & = -az +\pi i/2 + g_2(z), \qquad\qquad\quad\!\! z \in \mathcal R_3, \quad \Re z > 0, \\
    \lam_3(z) & = -az +\pi i/2 + g_1(z) - g_2(z), \quad\ \! z \in \mathcal R_3, \quad \Re z < 0.
\end{aligned}
\end{equation}
Note that $\lam_2(z)$ and $\lam_3(z)$ are defined with a cut along the entire
imaginary axis. In fact, from \eqref{eq:g2}--\eqref{def:lambda} it follows that
\begin{equation}\label{eq:cutlambdaimaginary}
\lam_{3,+}(z) - \lam_{3,-}(z) = \left\{\begin{array}{cl} 0,&\qquad z\in[ic,i\infty),\\
-\pi i,&\qquad z\in(-i\infty,-ic],
\end{array}\right.
\end{equation}
and a similar formula holds for $\lam_2$.

We can now reformulate Lemma~\ref{lemma:gfunctions} in terms of the
$\lam$-functions. This leads to the following two lemmas.

\begin{lemma} \label{lemma:lambda1} We have
\begin{equation}\label{eq:lambda1}
\begin{aligned}
    \lam_{2,\pm}(x)-\lam_{1,\mp}(x) & \left\{\begin{array}{ll} =0, \qquad x \in \mathbb R_+\cap\supp(\mu_1),\\
    < 0, \qquad x \in \mathbb R_+\setminus\supp(\mu_1),
    \end{array}\right.\\
    \lam_{3,\pm}(x)-\lam_{1,\mp}(x) & \left\{\begin{array}{ll}
    = 0, \qquad x \in \mathbb R_-\cap\supp(\mu_1),\\ < 0, \qquad x \in \mathbb R_-\setminus\supp(\mu_1),
    \end{array}\right.\\
    \Re(\lam_{3,+}(z)-\lam_{3,-}(z)) & \left\{\begin{array}{ll} = 0, \qquad z \in
    i\mathbb R\setminus(-ic,ic),\\< 0, \qquad z \in
    (-ic,ic).\end{array}\right.
\end{aligned}
\end{equation}
\end{lemma}

\begin{proof}
These are reformulations of the variational conditions
\eqref{eq:varcondition1}--\eqref{eq:varcondition4} associated with the
equilibrium problem, taking into account \eqref{eq:g1}--\eqref{def:lambda}. The
fact that we have strict inequalities follows from our assumption that the
measure $\mu_1$ is regular.
\end{proof}

\begin{lemma} \label{lemma:lambda2} We have
\begin{equation}\label{eq:lambda2}
\begin{aligned}
\lam_{1,+}(x)-\lam_{1,-}(x) & = -2\pi i\mu_1([x,+\infty)), \qquad x \in\mathbb R, \\
\lam_{2,+}(x)-\lam_{2,-}(x) & = 2\pi i\mu_1([x,+\infty)), \qquad x
\in\mathbb R_+, \\
\lam_{3,+}(x)-\lam_{3,-}(x) & =
2\pi i\mu_1([x,+\infty)), \qquad x \in\mathbb R_-,\\
\lam_{2,+}(z)-\lam_{3,-}(z) & = \lam_{2,-}(z)-\lam_{3,+}(z) = \\
& \left\{\begin{array}{ll} 0, \qquad\qquad\qquad\qquad\qquad\qquad\quad\ z \in (-ic,ic),\\
2\pi i\mu_2([z,i\infty))-\pi i/2+2az,\qquad z\in i\mathbb R.\end{array}\right.
\end{aligned}
\end{equation}
The last expression is purely imaginary and its imaginary part is strictly
increasing in terms of $\Im(z)$ as $z\in (-i\infty,-ic)\cup(ic,i\infty)$.
\end{lemma}

\begin{proof} This is a straightforward calculation using \eqref{eq:g1}--\eqref{def:lambda}.
\end{proof}

We define the new matrix valued function $T$ as
\begin{multline}\label{def:T}
    T(z) = \diag\left(e^{-n \ell}, 1, e^{-n \frac{1}{2} \pi i} \right) X(z) \\
    \times
    \diag \left( e^{n(\lam_1(z)-V(z))}, e^{n(\lam_2(z)-az)},
    e^{n(\lam_3(z)+az)}
    \right).
\end{multline}

Then $T$ satisfies the following RH problem.

\begin{itemize}
\item[(1)] $T$ is analytic on $\mathbb C \setminus (\mathbb R \cup [-ic,ic] \cup \Gamma)$.
\item[(2)] The jumps for $T$ are
\begin{align}
    T_+(x) & = T_-(x) \begin{pmatrix} e^{n(\lam_{1,+}-\lam_{1,-})} & e^{n(\lam_{2,+}-\lam_{1,-})} & 0 \\ 0 & e^{n(\lam_{2,+}-\lam_{2,-})} & 0 \\ 0 & 0 & 1 \end{pmatrix}
    \qquad \textrm{for } x > q, \\
    T_+(x) & = T_-(x) \begin{pmatrix} e^{n(\lam_{1,+}-\lam_{1,-})} & 0 & e^{n(\lam_{3,+}-\lam_{1,-})} \\ 0 & 1 & 0 \\ 0 & 0 & e^{n(\lam_{3,+}-\lam_{3,-})} \end{pmatrix}
    \qquad \textrm{for } x < -q,\\
    T_+(x) & = T_-(x) \begin{pmatrix} 1 & e^{n(\lam_{2}-\lam_{1})} & e^{n(\lam_{3}-\lam_{1})} \\ 0 & 1 & 0 \\ 0 & 0 & 1 \end{pmatrix}
    \qquad \textrm{for } -q< x < q,
\end{align}
on the real axis,
\begin{align}\label{jumpsTimag}
    T_+(z) & = T_-(z) \begin{pmatrix} 1 & 0 & 0 \\ 0 & 0 & 1 \\ 0 & -1 & e^{n(\lam_{3,+}-\lam_{3,-})} \end{pmatrix}
    \qquad \textrm{for } z \in (-ic,ic),
\end{align}
on the imaginary interval, and
\begin{align}
    \label{eq:jumpTGamma1}
    T_+(z) & = T_-(z) \begin{pmatrix} 1 & 0 & 0 \\ 0 & 1 & -e^{n(\lam_{3}-\lam_{2})} \\ 0 & 0 & 1 \end{pmatrix}
    \qquad \textrm{for } z\in\Gamma^+, \\
    \label{eq:jumpTGamma2}
    T_+(z) & = T_-(z) \begin{pmatrix} 1 & 0 & 0 \\ 0 & 1 & 0 \\ 0 & -e^{n(\lam_{2}-\lam_{3})} & 1 \end{pmatrix}
    \qquad \textrm{for } z\in\Gamma^-,
\end{align}
on the curve $\Gamma$.
\item[(3)]
As $z\to\infty$ we have that
\begin{equation*}
T(z) = I+O(1/z).
\end{equation*}
\end{itemize}

Note that for the $(2,3)$ and $(3,2)$ entries of the jump matrix in
\eqref{jumpsTimag}, we used the last formula in \eqref{eq:lambda2}. Also
observe that $T(z)$ does not have jumps on $i\mathbb R\setminus [-ic,ic]$.
Indeed, in principle one should consider the jump relation
\begin{equation}\label{redundantjump}
    T_+(z) = T_-(z) \begin{pmatrix} 1 & 0 & 0 \\ 0 & e^{n(\lam_{2,+}-\lam_{2,-})} & 0 \\ 0 & 0 & e^{n(\lam_{3,+}-\lam_{3,-})}
    \end{pmatrix},\qquad z\in(-i\infty,-ic)\cup (ic,i\infty).
\end{equation}
But from the jump relations in \eqref{eq:cutlambdaimaginary} we see that
$$e^{n(\lam_{3,+}(z)-\lam_{3,-}(z))} = e^{0} = 1,\qquad
z\in(ic,i\infty),$$ and also
$$e^{n(\lam_{3,+}(z)-\lam_{3,-}(z))} = e^{-n\pi i} = 1,\qquad
z\in(-i\infty,-ic),$$ by the fact that $n$ is even. In a similar way one shows
that $e^{n(\lam_{2,+}(z)-\lam_{2,-}(z))} = 1$, and so the jump matrix in
\eqref{redundantjump} is just the identity matrix.

By virtue of \eqref{eq:lambda1}--\eqref{eq:lambda2}, we may rewrite the jumps
on $\mathbb R$ as follows:
\begin{align} T_+ = T_-
    \begin{pmatrix}
    e^{n(\lam_{1,+} - \lam_{1,-})} & 1 & 0 \\
    0 & e^{-n(\lam_{1,+}-\lam_{1,-})} & 0 \\
    0 & 0 & 1 \end{pmatrix},
    \qquad x \in \supp(\mu_1), \, x > q, \end{align}
\begin{align} T_+ = T_-
    \begin{pmatrix}
    e^{n(\lam_{1,+} - \lam_{1,-})} & 0 & 1 \\
    0 & 1 & 0 \\
    0 & 0 & e^{-n(\lam_{1,+} - \lam_{1,-})} \end{pmatrix},
    \qquad x \in \supp(\mu_1), \, x < -q, \end{align}
and
\begin{align} T_+ = T_-
    \begin{pmatrix}
    e^{-2\pi i n \alpha_j} & e^{n(\lam_{2,+} - \lam_{1,-})} & 0 \\
    0 & e^{2\pi i n \alpha_j} & 0 \\
    0 & 0 & 1 \end{pmatrix},
    \qquad x \in (b_j, a_{j+1}), \, x > q, \end{align}
\begin{align} T_+ = T_-
    \begin{pmatrix}
    e^{-2\pi i n \alpha_j} & 0 & e^{n(\lam_{3,+} - \lam_{1,-})} \\
    0 & 1 & 0 \\
    0 & 0 & e^{2\pi i n \alpha_j}  \end{pmatrix},
    \qquad x \in (b_j, a_{j+1}), \, x < -q, \end{align}
where $b_0 = -\infty$, $a_{N+1} = + \infty$ and
\begin{align}
    \alpha_j = \mu_1([a_{j+1}, +\infty)), \qquad j=0, \ldots, N.
    \end{align}

We would like the jump matrices on $\Gamma$ to be exponentially close to the
identity matrix as $n\to\infty$. From the above RH problem, we see that this is
achieved provided $\Gamma$ lies in the region where $\Re(\lam_{3}-\lam_2)<0$ if
$\Re z>0$ and $\Re(\lam_{3}-\lam_2)>0$ if $\Re z<0$. The fact that $\Gamma$ can
indeed be chosen in this way, follows by applying the Cauchy-Riemann equations
to the last equality in Lemma~\ref{lemma:lambda2}, and using the last line in
the statement of that lemma.

\subsection{Lenses around the intervals $[a_j,b_j]$}
\label{subsection:lenses}

Around each of the intervals $[a_j,b_j]$ we open up a small lens to transform
the oscillatory entries of the jump matrix into exponentially decaying entries.
Since the non-trivial part of the jump matrix is locally of size $2\times 2$
only, this can be done in the standard way \cite{Dei}.

\begin{figure}[t]
\begin{center}
   \setlength{\unitlength}{1truemm}
   \begin{picture}(100,70)(-5,2)

       \put(0,40){\line(1,0){95}}
       \put(93,36.6){$\mathbb R $}
       \put(95,40){\thicklines\vector(1,0){.0001}}
       \put(28,40){\thicklines\circle*{1}}
       \put(27,36.6){$a_j$}
       \put(62,40){\thicklines\circle*{1}}
       \put(61,36.6){$b_j$}

       \qbezier(28,40)(45,60)(62,40)
       \put(45,50){\thicklines\vector(1,0){1}}
       \put(45,52){$\Gamma^+_j$}

       \qbezier(28,40)(45,20)(62,40)
       \put(45,30){\thicklines\vector(1,0){1}}
       \put(45,26){$\Gamma^-_j$}
   \end{picture}
 \vspace{-25mm}
   \caption{Lens around the interval $[a_j,b_j]$.}
   \label{fig:locallenses}
\end{center}
\end{figure}
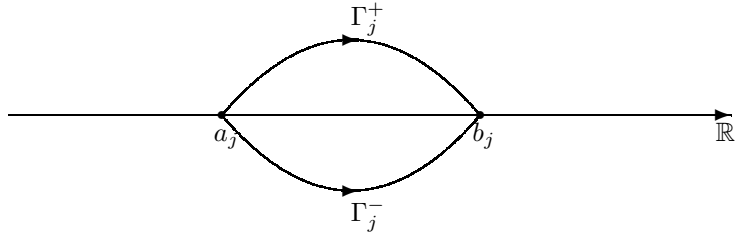

More precisely, we take Jordan curves $\Gamma_j^+$ and $\Gamma_j^-$ surrounding
the interval $[a_j,b_j]$ as in Figure~\ref{fig:locallenses}. The region between
these curves is called the lens, and $\Gamma_j^+$ and $\Gamma_j^-$ are the
upper and lower lip of the lens, respectively. We choose them sufficiently
close to the real axis so that
\begin{equation} \label{lenschoice1}\Re (\lam_{1}(z) - \lam_{2}(z) ) <
0,\qquad \textrm{ for }z \textrm{ inside the lens},\ z\not\in[a_j,b_j],\
\Re(z)>0,
\end{equation}
\begin{equation} \label{lenschoice2}\Re (\lam_{1}(z) - \lam_{3}(z) ) < 0,
\qquad \textrm{ for }z \textrm{ inside the lens},\ z\not\in[a_j,b_j],\
\Re(z)<0.
\end{equation}
The fact that this is possible, follows from applying the Cauchy-Riemann
equations to the first equation of \eqref{eq:lambda2}, cf.~\cite{Dei}.

The curves $\Gamma_{j}^{\pm}$ are chosen disjoint from each other, disjoint
from the lens $\Gamma$ in Section~\ref{subsection:transfo1}, and also disjoint
from the imaginary axis. The only case where an intersection with the imaginary
axis occurs is in Case~III with $j=(N+1)/2$, and we assume that in this case,
$\Gamma_{j}^{\pm}$ intersects $i\mathbb R$ at a point $\pm iz_0$ strictly
inside $(-ic,ic)$.

We define
\begin{equation}\label{def:S}
S(z) = \left\{\begin{array}{l} T(z)\begin{pmatrix} 1 & 0 & 0 \\
-e^{n (\lam_{1}(z)-\lam_{2}(z))} & 1 & 0 \\ 0 & 0 & 1
\end{pmatrix},\qquad\!\!\!\! z\textrm{ in upper part lens}, \Re z > 0, \\
T(z)\begin{pmatrix} 1 & 0 & 0 \\ e^{n (\lam_{1}(z)-\lam_{2}(z))} & 1 & 0
\\ 0 & 0 & 1
\end{pmatrix},\qquad z\textrm{ in lower part lens}, \Re z > 0, \\
T(z)\begin{pmatrix} 1 & 0 & 0 \\ 0 & 1 & 0 \\
-e^{n (\lam_{1}(z)-\lam_{3}(z))} &  0 & 1
\end{pmatrix},\qquad\!\!\!\! z\textrm{ in upper part lens}, \Re z < 0, \\
T(z)\begin{pmatrix} 1 & 0 & 0 \\ 0 & 1 & 0  \\
e^{n (\lam_{1}(z)-\lam_{3}(z))} & 0 & 1
\end{pmatrix},\qquad z\textrm{ in lower part lens}, \Re z < 0, \\
T(z),\qquad\qquad\qquad\qquad\qquad\qquad \textrm{ elsewhere}.
\end{array}\right.
\end{equation}

Then $S$ satisfies the following RH problem.

\begin{itemize}
\item[\rm (1)] $S$ is analytic
in $\mathbb C \setminus\left(\mathbb R\cup i\mathbb R\cup\Gamma\cup\bigcup_{j}
\Gamma_j )\right)$.
 \item[\rm (2)] For $x \in \supp(\mu_1)$ we have
that
\begin{equation*}\label{jumpsS1}
S_{+}(x) = S_{-}(x)
\begin{pmatrix}
0 & 1 & 0 \\ -1 & 0 & 0 \\ 0 & 0 & 1
\end{pmatrix},\qquad x\in \supp(\mu_1)\cap\mathbb R_+,
\end{equation*}
\begin{equation*}\label{jumpsS2}
S_{+}(x) = S_{-}(x)
\begin{pmatrix}
0 & 0 & 1 \\ 0 & 1 & 0 \\ -1 & 0 & 0
\end{pmatrix},\qquad x\in \supp(\mu_1)\cap\mathbb R_-.
\end{equation*}
On the lips $\Gamma_j$ of the lenses we have
\begin{equation*} \label{jumpsS3}
S_{+}(z) = S_{-}(z)
\begin{pmatrix} 1 & 0 & 0 \\ e^{n(\lam_{1}-\lam_{2})} & 1 & 0 \\ 0 & 0 & 1\end{pmatrix},
\qquad \text{for }  z\in \Gamma^{\pm}_j, \quad \Re z>0,
\end{equation*}
\begin{equation*} \label{jumpsS4}
S_{+}(z) = S_{-}(z)
\begin{pmatrix} 1 & 0 & 0 \\ 0 & 1 & 0 \\ e^{n(\lam_{1}-\lam_{3})} & 0 & 1\end{pmatrix},
\qquad \text{for }  z\in \Gamma^{\pm}_j, \quad \Re z <0.
\end{equation*}
In Case III we also have
\begin{equation}\label{jumpSredundant}
    S_{+}(z) = S_{-}(z)
\qquad \textrm{for } z \in (-iz_0,iz_0).
\end{equation}

The jumps of $S(z)$ on the other contours are the same as those for $T(z)$.
\item[\rm (3)] As $z\to\infty$, we have that
\begin{equation*}\label{asymptoticconditionS}
    S(z) =I+O(1/z).
\end{equation*}
\end{itemize}

From \eqref{eq:lambda1} and \eqref{lenschoice1}--\eqref{lenschoice2}, it can be
checked that all the non-constant entries in the jump matrices for $S(z)$ tend
to $0$ as $n \to \infty$, uniformly for $z$ bounded away from the branch points
$a_j$, $b_j$, $\pm ic$. The only case that requires more explanation is the
$(3,1)$ entry in the jump matrix in \eqref{jumpSredundant}. In that case, one
can factorize
$$ e^{n (\lam_{1}(z)-\lam_{3,-}(z))} = e^{n (\lam_{1}(z)-\lam_{3,+}(z))}e^{n
(\lam_{3,+}(z)-\lam_{3,-}(z))},
$$
and observe from \eqref{eq:lambda1} and \eqref{lenschoice2} that for
$z\in[-iz_0,iz_0]$, the leftmost factor is uniformly bounded by $1$ while the
rightmost factor is uniformly exponentially decaying as $n\to\infty$.

\subsection{Global parametrix}
\label{subsection:global}

The global parametrix we look for is a $3\times 3$ matrix valued function $M$
with jumps (obtained from the jumps of $S$ by ignoring all entries which are
exponentially small for $n\to\infty$)

\begin{itemize}
\item[\rm (1)] $M$ is analytic in $\mathbb C \setminus ([a_1,b_N] \cup [-ic,ic])$,
\item[\rm (2)] The jumps for $M$ are
\begin{align}
    M_+(x) & = M_-(x)
    \begin{pmatrix} 0 & 1 & 0 \\ -1 & 0 & 0 \\ 0 & 0 & 1 \end{pmatrix},\qquad x\in\supp(\mu_1) \cap \mathbb R_+, \\
    M_+(x) & = M_-(x)
    \begin{pmatrix} 0 & 0 & 1 \\ 0 & 1 & 0  \\ -1 & 0 & 0 \end{pmatrix},\qquad x\in\supp(\mu_1) \cap \mathbb
    R_-, \\
    M_+(x) & = M_-(x) \begin{pmatrix} e^{-2\pi i n \alpha_j} & 0 & 0 \\
    0 & e^{2\pi i n \alpha_j} & 0 \\ 0 & 0 & 1 \end{pmatrix},
        \qquad x\in(b_j, a_{j+1}) \cap \mathbb R_+, \\
    M_+(x) & = M_-(x) \begin{pmatrix} e^{-2\pi i n \alpha_j} & 0 & 0  \\
    0 & 1 & 0 \\0 & 0 & e^{2\pi i n \alpha_j} \end{pmatrix},
        \qquad  x\in(b_j, a_{j+1}) \cap \mathbb R_-,
\end{align}
on the real line, and
\begin{align}
    M_+(z) & = M_-(z) \begin{pmatrix} 1 & 0 & 0 \\ 0 & 0 & 1 \\ 0 & -1 & 0
        \end{pmatrix},
    \quad z \in (-ic,ic). \end{align}
\item[\rm (3)] At infinity we have
\[ M(z) = I + O(1/z), \qquad \text{as } z \to\infty, \]
\item[\rm (4)]  $M$ has at most fourth-root singularities at the branch points $a_1,b_1,
\ldots, a_N$, $b_N$, $\pm ic$.
\end{itemize}

\begin{figure}[t]
\begin{center}
   \setlength{\unitlength}{1truemm}
   \begin{picture}(100,50)(-5,2)

       \put(7,29){\line(1,0){55}}
       \put(19,41){\line(1,0){55}}
       \put(7,29){\line(1,1){12}}
       \put(62,29){\line(1,1){12}}
       \thicklines
       \put(20,35){\line(1,0){8}}
       \put(32,35){\line(1,0){7}}
       \put(44,35){\line(1,0){7}}
       \put(55,35){\line(1,0){8}}
       \thinlines
       \put(75,35){$\mathcal R_1$}
       \qbezier(16,35)(17,39)(33.5,40)
       \qbezier(33.5,40)(52,39)(53,35)
       \qbezier(16,35)(17,31)(33.5,30)
       \qbezier(33,30)(52,31)(53,35)

       \qbezier(18,35)(19,37.5)(24,38)
       \qbezier(24,38)(29,37.5)(30,35)
       \qbezier(18,35)(19,32.5)(24,32)
       \qbezier(24,32)(29,32.5)(30,35)
       \qbezier(25,35)(26,37.5)(30,38)
       \qbezier(30,38)(34,37.5)(35,35)
       \qbezier(48,35)(49,37.5)(53,38)
       \qbezier(53,38)(57,37.5)(58,35)
       \put(31,42){$\small{B_2}$}
       \put(21,40){$\small{B_1}$}
       \put(34,40){\thicklines\vector(-1,0){1}}
       \put(24,38){\thicklines\vector(-1,0){1}}

       \put(7,14){\line(1,0){55}}
       \put(19,26){\line(1,0){55}}
       \put(7,14){\line(1,1){12}}
       \put(62,14){\line(1,1){12}}
       \thicklines
       \put(44,20){\line(1,0){7}}
       \put(55,20){\line(1,0){8}}
       \thinlines
       \qbezier(48,20)(49,17.5)(53,17)
       \qbezier(53,17)(57,17.5)(58,20)
       \put(53,17){\thicklines\vector(-1,0){1}}
       \put(50,14){$\small{A_2}$}
       \put(75,20){$\mathcal R_2$}

       \put(7,-1){\line(1,0){55}}
       \put(19,11){\line(1,0){55}}
       \put(7,-1){\line(1,1){12}}
       \put(62,-1){\line(1,1){12}}
       \thicklines
       \put(20,5){\line(1,0){8}}
       \put(32,5){\line(1,0){7}}
       \thinlines
       \qbezier(25,5)(26,2.5)(30,2)
       \qbezier(30,2)(34,2.5)(35,5)

       \put(30,2){\thicklines\vector(-1,0){1}}
       \put(27,-1){$\small{A_1}$}
       \put(75,5){$\mathcal R_3$}

   \end{picture}
   \vspace{1mm}
   \caption{Canonical homology basis in Case I (with $N=4$).}
   \label{fig:homologybasis1}
\end{center}
\end{figure}
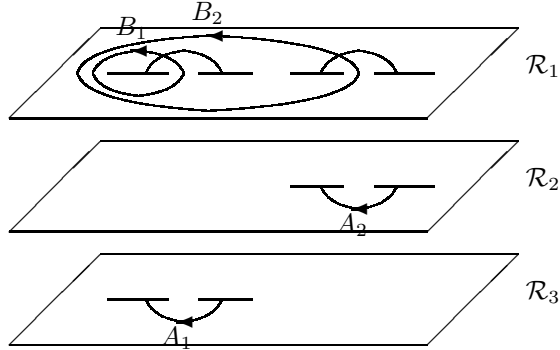

We can solve this problem with the help of meromorphic
differentials on the Riemann surface. Such a construction was
first used in \cite{Mo} and later developed further in
\cite{DuKuMo,KMo}.

\begin{figure}[t]
\begin{center}
   \setlength{\unitlength}{1truemm}
   \begin{picture}(100,50)(-5,2)

       \put(7,29){\line(1,0){55}}
       \put(19,41){\line(1,0){55}}
       \put(7,29){\line(1,1){12}}
       \put(62,29){\line(1,1){12}}
       \thicklines
       \put(20,35){\line(1,0){8}}
       \put(32,35){\line(1,0){7}}
       \put(44,35){\line(1,0){7}}
       \put(55,35){\line(1,0){8}}
       \thinlines
       \put(75,35){$\mathcal R_1$}
       \qbezier(16,35)(17,39)(33.5,40)
       \qbezier(33.5,40)(52,39)(53,35)
       \qbezier(16,35)(17,31)(33.5,30)
       \qbezier(33,30)(52,31)(53,35)

       \qbezier(17,35)(18,31.5)(29,31)
       \qbezier(29,31)(40,31.5)(41,35)
       \qbezier(17,35)(18,38.5)(29,39)
       \qbezier(29,39)(40,38.5)(41,35)

       \qbezier(18,35)(19,37.5)(24,38)
       \qbezier(24,38)(29,37.5)(30,35)
       \qbezier(18,35)(19,32.5)(24,32)
       \qbezier(24,32)(29,32.5)(30,35)
       \qbezier(25,35)(26,37.5)(30,38)
       \qbezier(30,38)(34,37.5)(35,35)
       \qbezier(37,35)(38,37.5)(41.5,38)
       \qbezier(41.5,38)(45,37.5)(46,35)
       \qbezier(48,35)(49,37.5)(53,38)
       \qbezier(53,38)(57,37.5)(58,35)
       \put(31,42){$\small{B_3}$}
       \put(26,41){$\small{B_2}$}
       \put(21,40){$\small{B_1}$}
       \put(34,40){\thicklines\vector(-1,0){1}}
       \put(29,39){\thicklines\vector(-1,0){1}}
       \put(24,38){\thicklines\vector(-1,0){1}}

       \put(7,14){\line(1,0){55}}
       \put(19,26){\line(1,0){55}}
       \put(7,14){\line(1,1){12}}
       \put(62,14){\line(1,1){12}}
       \thicklines
       \put(44,20){\line(1,0){7}}
       \put(55,20){\line(1,0){8}}
       \put(38.5,17){\line(1,1){6}}
       \thinlines
       \qbezier(48,20)(49,17.5)(53,17)
       \qbezier(53,17)(57,17.5)(58,20)
       \qbezier(41.5,17)(45,17.5)(46,20)
       \qbezier(39.6,17.9)(40,17.3)(41.5,17)
       \put(41.5,17){\thicklines\vector(-1,0){1}}
       \put(53,17){\thicklines\vector(-1,0){1}}
       \put(38.5,14){$\small{A_2}$}
       \put(50,14){$\small{A_3}$}
       \put(75,20){$\mathcal R_2$}

       \put(7,-1){\line(1,0){55}}
       \put(19,11){\line(1,0){55}}
       \put(7,-1){\line(1,1){12}}
       \put(62,-1){\line(1,1){12}}
       \thicklines
       \put(20,5){\line(1,0){8}}
       \put(32,5){\line(1,0){7}}
       \put(38.5,2){\line(1,1){6}}
       \thinlines
       \qbezier(25,5)(26,2.5)(30,2)
       \qbezier(30,2)(34,2.5)(35,5)
       \qbezier(37,5)(37.5,3.5)(39.5,3)

       \put(30,2){\thicklines\vector(-1,0){1}}
       \put(27,-1){$\small{A_1}$}
       \put(75,5){$\mathcal R_3$}

   \end{picture}
   \vspace{1mm}
   \caption{Canonical homology basis in Case II (with $N=4$).}
   \label{fig:homologybasis2}
\end{center}
\end{figure}
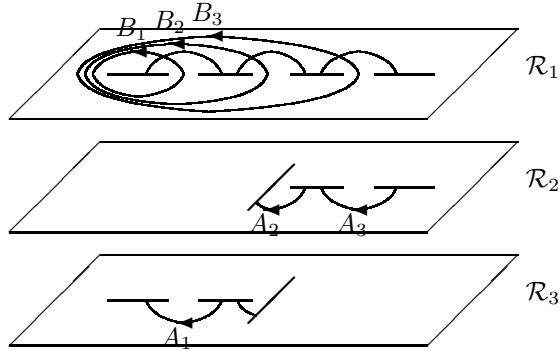

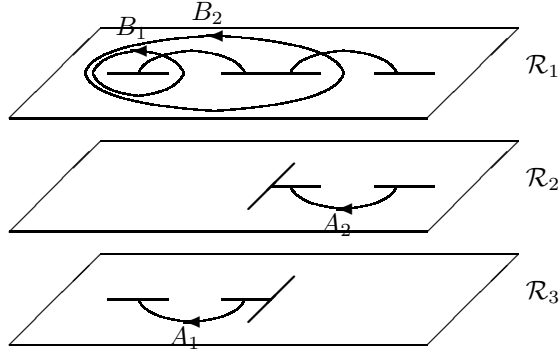
\begin{figure}[t]
\begin{center}
   \setlength{\unitlength}{1truemm}
   \begin{picture}(100,50)(-5,2)

       \put(7,29){\line(1,0){55}}
       \put(19,41){\line(1,0){55}}
       \put(7,29){\line(1,1){12}}
       \put(62,29){\line(1,1){12}}
       \thicklines
       \put(20,35){\line(1,0){8}}
       \put(35,35){\line(1,0){13}}
       \put(55,35){\line(1,0){8}}
       \thinlines
       \put(75,35){$\mathcal R_1$}
       \qbezier(17,35)(18,39)(34,40)
       \qbezier(34,40)(50,39)(51,35)
       \qbezier(17,35)(18,31)(34,30)
       \qbezier(34,30)(50,31)(51,35)
       \qbezier(18,35)(19,37.5)(24,38)
       \qbezier(24,38)(29,37.5)(30,35)
       \qbezier(18,35)(19,32.5)(24,32)
       \qbezier(24,32)(29,32.5)(30,35)
       \qbezier(24,35)(25,37.5)(31,38)
       \qbezier(31,38)(37,37.5)(38,35)
       \qbezier(44,35)(45,37.5)(51,38)
       \qbezier(51,38)(57,37.5)(58,35)
       \put(31,42){$\small{B_2}$}
       \put(21,40){$\small{B_1}$}
       \put(34,40){\thicklines\vector(-1,0){1}}
       \put(24,38){\thicklines\vector(-1,0){1}}

       \put(7,14){\line(1,0){55}}
       \put(19,26){\line(1,0){55}}
       \put(7,14){\line(1,1){12}}
       \put(62,14){\line(1,1){12}}
       \thicklines
       \put(41.5,20){\line(1,0){6.5}}
       \put(55,20){\line(1,0){8}}
       \put(38.5,17){\line(1,1){6}}
       \thinlines
       \qbezier(44,20)(45,17.5)(51,17)
       \qbezier(51,17)(57,17.5)(58,20)
       \put(51,17){\thicklines\vector(-1,0){1}}
       \put(48,14){$\small{A_2}$}
       \put(75,20){$\mathcal R_2$}

       \put(7,-1){\line(1,0){55}}
       \put(19,11){\line(1,0){55}}
       \put(7,-1){\line(1,1){12}}
       \put(62,-1){\line(1,1){12}}
       \thicklines
       \put(20,5){\line(1,0){8}}
       \put(35,5){\line(1,0){6.5}}
       \put(38.5,2){\line(1,1){6}}
       \thinlines
       \qbezier(24,5)(25,2.5)(31,2)
       \qbezier(31,2)(37,2.5)(38,5)
       \put(31,2){\thicklines\vector(-1,0){1}}
       \put(28,-1){$\small{A_1}$}
       \put(75,5){$\mathcal R_3$}

   \end{picture}
   \vspace{1mm}
   \caption{Canonical homology basis in Case III (with $N=3$).}
   \label{fig:homologybasis3}
\end{center}
\end{figure}

To the Riemann surface we associate a canonical homology basis $\{A_1, \ldots,
A_g$, $B_1, \ldots, B_g\}$ where $g$ is the genus. The details of the construction
depend on whether we are in Case I, II or III, see
Figures~\ref{fig:homologybasis1}--\ref{fig:homologybasis3}.

For brevity, we give a detailed description only for Cases II and III. Then the
genus is $g = N -1$. The cycles $B_j$ are on the first sheet and $B_j$
encircles $[a_1,b_j]$ once in the counterclockwise direction. The cycles $A_j$
are partly in the upper half-plane on the first sheet and partly in the lower
half-plane on the second or third sheet. $A_j$ passes through $[a_j,b_j]$ and
$[a_{j+1},b_{j+1}]$.

The anti-holomorphic involution $\phi$ is defined by mapping $z$ to
$\overline{z}$ on the same sheet. The fixed point set of $\phi$ is the
disjoint union of $N = g+1$ closed curves $\bigcup_{j=0}^{g} \Sigma_j$ on the
Riemann surface. Here $\Sigma_j$ is homotopic to $A_j$ as a closed curve,
$j=1,\ldots,g$, while $\Sigma_0$ is the unbounded component.

\begin{lemma}
If $P_j \in \Sigma_j$ for $j=1, \ldots, g$, then the divisor
\[ D = \sum_{j=1}^g P_j \]
is non-special.
\end{lemma}

\begin{proof}
A detailed proof of this theorem in a similar setting will be
given in \cite{DuKuMo}. Here we only outline some of the key
steps. We use $u: \mathcal R \to \mathbb C^g/L$ to denote the Abel
map, mapping the Riemann surface $\mathcal R$ to $\mathbb C^g/L$,
where $L$ is the lattice defined from the columns of the Riemann
period matrix. We also use $\theta : \mathbb C^g \to \mathbb C^g$
to denote the corresponding Riemann theta function on $\mathbb
C^g$, and $\vec{K}$ denotes the vector of Riemann constants, see
e.g.\ \cite{FK}.

The proof is based on the following result which can be found e.g.\ in
\cite[Theorem 2.4.2]{Dubr}. The divisor $D$ is non-special if and only if
$\theta(u(P)-u(D)-\vec{K})$ does not vanish identically for $P$ on the Riemann
surface.

The next step is to show that for $\vec{s}\in\mathbb R^g$ we have
$\theta(\vec{s})>0$. This relies on the fact that the Riemann surface
$\mathcal R$ is the Schottky double of a bordered Riemann surface, see
\cite[Corollary 6.13]{Fay} and \cite{Mo}.

Using the antiholomorphic involution $\phi$ it can be shown that
the Riemann period matrix is purely imaginary. Finally, one then
shows that $u(D)+\vec{K}$ has a real representative modulo the
lattice $L$ \cite{DuKuMo}. By taking into account the results in
the last two paragraphs, the desired result then follows.
\end{proof}

We now basically follow \cite{KMo}. Given $(P_1, \ldots, P_g)$ with $P_j \in
\Sigma_j$, we define a meromorphic differential $\omega_P$ so that
\begin{itemize}
\item $\omega_P$ has simple poles in $a_1,b_1,\ldots, a_N, b_N$, $\pm ic$,
$P_1, \ldots, P_g$, $\infty_2$ and $\infty_3$ with residues
\begin{align*}
    \Res(\omega_P, a_j)  & = \Res(\omega_P, b_j) = -\tfrac{1}{2},
        \qquad j=1,\ldots, N, \\
    \Res(\omega_P, \pm ic) & = - \tfrac{1}{2}, \\
    \Res(\omega_P, P_j) & = 1,   \qquad j=1, \ldots, g, \\
    \Res(\omega_P, \infty_2) & = \Res(\omega_P, \infty_3) = 1,
    \end{align*}
\item $\omega_P$ is holomorphic elsewhere,
\item $\omega_P$ has vanishing $A$-periods:
\begin{align*}
    \int_{A_j} \omega_P = 0, \qquad j=1, \ldots, g.
    \end{align*}
\end{itemize}

The total sum of the residues is $0$. These conditions determine $\omega_P$
uniquely. The $B$-periods are purely imaginary, they are well defined modulo
$2\pi i \mathbb Z$, and the mapping
\begin{multline*}
    \Psi : \Sigma_1 \times \cdots \times \Sigma_g \to (\mathbb R \slash \mathbb
    Z)^g:
    (P_1, \ldots, P_g) \mapsto \frac{1}{2\pi i}
        \left(\int_{B_1} \omega_P, \ldots, \int_{B_g} \omega_P \right) \end{multline*}
is a bijection. These claims follow in the same way as in \cite{KMo}.

Thus there exist $P_j^{(1)} \in \Sigma_j$ so that
\[ \Psi(P_1^{(1)}, \ldots, P_g^{(1)}) = (n \alpha_1, \ldots, n \alpha_g)
    \qquad \mod \mathbb Z. \]
Let $\omega_P^{(1)}$ be the corresponding meromorphic differential.

We take the base point $P_0 =\infty_1$ and define three functions
$v_1(z)$, $v_2(z)$ and $v_3(z)$ of a complex variable $z$ as follows.
We have
\[ v_k(z) = \exp \left(\int_{P_0}^z \omega_P^{(1)} \right) \]
where $z$ is considered as a point on the $k$th sheet of $\mathcal R$, and
where the path of integration is as follows
\begin{itemize}
\item for $k=1$, the path of integration is on the first sheet
and  does not intersect the real line,
\item for $k=2$, the path of integration is on the first and second sheets;
for $\Im z > 0$ ($\Im z < 0$), the path starts in the lower (upper) half-plane
of the first sheet, crosses one of the cuts $[a_j,b_j]$ in $\mathbb R_+$ to the
upper (lower) half-plane of the second sheet and stays there until it ends at
$z$;
\item for $k=3$, the path of integration is on the first and third sheets;
for $\Im z > 0$ ($\Im z < 0$), the path starts in the lower (upper) half-plane
of the first sheet, crosses one of the cuts $[a_j,b_j]$ in $\mathbb R_-$ to the
upper (lower) half-plane of the third sheet and stays there until it ends at
$z$.
\end{itemize}

Then the vector $(v_1, v_2, v_3)$ is well-defined and analytic in $\mathbb C \setminus (\mathbb R \cup i \mathbb R)$
with jumps
\[ (v_1, v_2, v_3)_+ = (v_1, v_2, v_3)_- J_v \]
where the jump matrices on the real line are
\[ J_v(x)  = \begin{pmatrix} 0 & 1 & 0 \\ 1 & 0 & 0 \\ 0 & 0 & -1 \end{pmatrix},
    \qquad \text{for } x \in \supp(\mu_1)\cap\mathbb R_+, \]
\[ J_v(x)  = \begin{pmatrix} 0 & 0 & 1 \\ 0 & -1 & 0 \\ 1 & 0 & 0 \end{pmatrix},
    \qquad \text{for } x \in \supp(\mu_1)\cap\mathbb R_-, \]
\[ J_v(x) = \begin{pmatrix} 1 & 0 & 0 \\ 0 & -1 & 0 \\ 0 & 0 & -1 \end{pmatrix},
    \qquad \text{for } x < a_1 \text{ and } x > b_N, \]
\[ J_v(x) = \begin{pmatrix} e^{-2\pi i n \alpha_j} & 0 & 0 \\
        0 & - e^{2\pi in \alpha_j} & 0 \\ 0 & 0 & -1 \end{pmatrix}, \qquad
    \text{for } b_j <  x < a_{j+1},\quad x>0, \]
\[ J_v(x) = \begin{pmatrix} e^{-2\pi i n \alpha_j} & 0 & 0 \\
        0 & -1 & 0 \\ 0 & 0 & - e^{2\pi in \alpha_j} \end{pmatrix}, \qquad
    \text{for } b_j <  x < a_{j+1},\quad x<0,\]
while on the imaginary axis they are
\begin{equation*}
\begin{array}{ll}
J_v(z) = \begin{pmatrix} 1 & 0 & 0 \\ 0 & 1 & 0 \\ 0 & 0 & 1 \end{pmatrix},
    & \qquad \text{for } z \in (-i \infty, -ic) \cup (ic, i\infty),\\
J_v(z) = \begin{pmatrix} 1 & 0 & 0 \\ 0 & 0 & 1 \\ 0 & -1 & 0 \end{pmatrix},
    & \qquad \text{for } z \in (-ic, ic).
\end{array}
\end{equation*}

Also as $z \to \infty$, we have
\[ v_1(z) = 1 + O(1/z), \qquad v_2(z) = O(1/z), \qquad v_3(z) = O(1/z), \]
and all $v_j$ functions have fourth-root singular behavior at the branch
points $a_1, b_1, \ldots, a_N,b_N$, $\pm ic$.

Now define the first row $(M_{11}, M_{12}, M_{13})$ as follows
\begin{align*}
    \left(M_{11}, M_{12}, M_{13} \right)
    &= \left(v_1, v_2, v_3 \right) \qquad \text{if } \Im z > 0, \\
    &= \left(v_1,-v_2, -v_3 \right) \qquad \text{if } \Im z < 0.
    \end{align*}
This gives the correct jumps for $M$. The other rows of $M$ can be constructed
in a similar way, or by a simple transformation of the first row \cite{KMo}.

\subsection{Local parametrices}
\label{subsection:local}

In the regular case (the one we are considering) we construct local
parametrices out of Airy functions near each of the branch points. We denote
the local parametrices by $P$. The local parametrices match with the global
parametrix on the boundary of a small circle around the branch points. Since
the non-trivial part of the jump matrix is locally of size $2\times 2$ only,
and since in the regular case we have square root behavior near all the branch
points (Lemma~\ref{prop:squareroot}), this construction can be done in the
usual way \cite{Dei}. We omit the details.

\subsection{Final RH problem}
\label{subsection:transfofinal}

We define the final RH matrix \begin{equation*} R(z) = \left\{\begin{array}{ll}
T(z) M(z)^{-1} & \qquad \textrm{away from the branch points,} \\
T(z) P(z)^{-1} & \qquad \textrm{near the branch points.}
\end{array}\right.\end{equation*}
Then $R(z)$ has jump matrices tending to the identity matrix as $n\to\infty$
and it is normalized in the sense that $R(z)=I+O(1/z)$ as $z\to\infty$.
Standard arguments then show that $R(z)$ itself also tends to the identity
matrix, uniformly for $z$ on compact subsets of $\mathbb C$ \cite{Dei}.

\subsection{Proof of Theorem~\ref{theorem:main}}
\label{subsection:proofmaintheorem}

Having performed the steepest descent analysis of the RH problem, we can now
prove Theorem~\ref{theorem:main}. The proof follows the same pattern as in the
papers \cite{ABK,BK2,BK3,DelKui1}.

First assume that $x,y\in (a_j,b_j)$ with $x,y>0$. We will transform
\eqref{eq:KninY} under the series of transformations $Y\mapsto X\mapsto
T\mapsto S$. By virtue of \eqref{def:X} we have
\begin{equation*} \label{correlationkernelX}
    K_n(x,y) = \frac{1}{2\pi i(x-y)}
    \begin{pmatrix} 0 & w_1(y) & 0 \end{pmatrix}
    X_+(y)^{-1} X_+(x) \begin{pmatrix} 1 \\ 0 \\ 0 \end{pmatrix}.
    \end{equation*}
Using \eqref{def:T} this becomes
\begin{equation*} \label{correlationkernelT}
    K_n(x,y) = \frac{e^{n(V(x)-V(y))}}{2\pi i(x-y)}
    \begin{pmatrix} 0 & e^{n\lam_{2,+}(y)} & 0 \end{pmatrix}
    T_+(y)^{-1} T_+(x) \begin{pmatrix} e^{-n\lam_{1,+}(x)} \\ 0 \\ 0 \end{pmatrix}.
    \end{equation*}
From \eqref{def:S} we get
\begin{equation} \label{correlationkernelS}
    K_n(x,y) = \frac{e^{n(V(x)-V(y))}}{2\pi i(x-y)}
    \begin{pmatrix} -e^{n\lam_{1,+}(y)} & e^{n\lam_{2,+}(y)} & 0 \end{pmatrix}
    S_+(y)^{-1} S_+(x) \begin{pmatrix} e^{-n\lam_{1,+}(x)} \\ e^{-n\lam_{2,+}(x)} \\ 0 \end{pmatrix}.
    \end{equation}
Now it follows by standard arguments (e.g.\ \cite[Section 9]{BK2}) that
\begin{equation*}
S^{-1}(y)S(x) = I+O(x-y),\quad \textrm{as }y\to x
\end{equation*}
uniformly in $n$. Inserting this into \eqref{correlationkernelS} and setting
$h(x)=V(x)-\Re(\lam_1(x))$ yields
\begin{multline*} \label{correlationkernelS2}
    K_n(x,y)= e^{n(h(x)-h(y))}
    \frac{\sin(n\Im(\lam_{1,+}(x)-\lam_{1,+}(y)))}{\pi
    (x-y)}+O(1),\quad y\to x,
\end{multline*}
uniformly in $n$. By letting $y \to x$ and using l'H\^opital's rule we find
\begin{align*}K_n(x,x) &= \frac{n\Im(\xi_{1,+})}{\pi}+O(1)
\end{align*}
as $n \to \infty$. From \eqref{def:xi} and the Stieltjes inversion principle we
conclude that
\begin{equation*}
\lim_{n\to\infty} \frac{1}{n} K_n(x,x) = \frac{d\mu_1(x)}{dx}.
\end{equation*}
This proves \eqref{eq:Knlimit} for $x\in(a_j,b_j)$, $x>0$. The proof for the
other values of $x\in\mathbb R$ is similar, or can be obtained from symmetry
considerations.

\section{Phase transitions: General discussion}
\label{section:phasetransitions}

Recall the Cases I, II and III describing the structure of the equilibrium
measures in Section~\ref{subsection:threecases}. Intuitively one expects the
following possible behavior in terms of the parameter $a$. For large $a$ we are
in Case I. The measure $\mu_1$ is supported on two (or more) disjoint intervals
with a gap around $0$. The constraint $\sigma$ is not active.

When $a$ decreases the gap around $0$ shrinks. Then one of two things could
happen. It could happen that for a certain value of $a$ the constraint becomes
active, while the gap in the support of $\mu_1$ around $0$ is still there. Then
we are in Case II. Then if $a$ further decreases the gap may close or not. The
latter depends on whether in the unitary matrix model with potential $V$
(without external source) $0$ is in the support or not. If the support closes
then we are in Case III.

The other situation that could happen is that the constraint $\sigma$ remains
inactive all the way until for a certain value of $a$ the gap in the support of
$\mu_1$ is closed. Then if $a$ further decreases the constraint becomes active.
The transition is then from Case I to Case III without passing through the Case
II. This is precisely what happens in the quadratic case $V(x) =
\frac{1}{2}x^2$. More generally, one expects this kind of behavior when the
potential $V(x)$ is convex, or \lq nearly\rq\ convex.

For those values of $a$ for which a transition between one of the Cases I, II,
III to another takes places, one expects that the local eigenvalue correlations
near the origin are described by special functions related to ODE's. For
typical cases one expects such special functions as Pearcey integrals
\cite{AvM1,BK3,TW1} and the Hastings-McLeod solution to the Painlev\'e II
equation \cite{BI2,CK}. However, our model
allows for new kinds of critical and multi-critical behavior as well,
but it remains an open problem to describe these new critical phenomena.

In Section \ref{section:quartic} we will illustrate the above considerations in
detail for the case of a quartic potential. See in particular
Figure~\ref{fig:PhaseDiagram}.

\section{A case study: The quartic potential}
\label{section:quartic}

Let us investigate the case of a quartic potential
\[ V(z) = \frac{1}{4} z^4 - \frac{t}{2} z^2 \]
and the associated  McLaughlin equation
\eqref{eq:McLaughlin}:
\begin{equation}\label{eq:McLbis} \xi^3 - \left(z^3 -
tz\right) \xi^2 + (z^2 + \alpha) \xi + a^2 z^3 + \beta z = 0.
\end{equation} The discriminant of the McLaughlin equation (w.r.t.\ $\xi$) is a
polynomial $D_{12}(z)$ of degree $12$ in $z$. We calculated it with Maple, but
it is too long and not too interesting to reproduce it here in full. The first
terms are
\begin{equation} \label{eq:D12}
    D_{12}(z) = -4\alpha^3 + (\alpha^2 t^2 + 18\alpha \beta t - 12 \alpha^2 -27 \beta^2) z^2 +  O(z^4)
    \qquad \textrm{as } z \to 0.
    \end{equation}

The branch points of the Riemann surface are among the zeros
of $D_{12}(z)$. There are other zeros, and they should
come with higher multiplicities.

For general $\alpha$ and $\beta$ the McLaughlin equation has genus $4$
(according to Maple). The special choices for $\alpha$ and $\beta$ that are
relevant to us will lead to a reduction of the genus. The genus can be at most
one, as the following lemma shows.

A similar result occurs in \cite[Prop.~5.2.5]{DuK}, but the
proof given there is incorrect. Here we give a self-contained proof which may
be used for the situation in \cite{DuK} as well. The proof uses an idea due to
Lun Zhang (personal communication).

\begin{lemma} \label{lem:Vconvex}
Assume that $x \mapsto V(\sqrt{x})$ is convex for $x > 0$. Then the support of $\mu_1$
is either one interval (in case $0 \in \supp(\mu_1)$) or a disjoint union of
two intervals (in case $0 \not\in\supp(\mu_1)$), and the measure $\mu_1$ can
have singular behavior only at zero.
\end{lemma}

\begin{proof}
By fixing $\mu_2 \leq \sigma$ in the energy functional \eqref{eq:energyfunctional},
we see that $\mu_1$ is the unique minimizer for
the energy functional
\[
I(\mu)+\int_{-\infty}^{\infty}\left(V(x)-a|x|-U^{\mu_2}(x)\right) d\mu(x)
\]
among probability measures $\mu$ on $\mathbb R$.
Since the external field is symmetric, it follows from \cite[Theorem
IV.1.10(f)]{SaffTotik} that $d\mu_1(t)=d\tilde{\mu}_1(t^2)/2$ where $\tilde{\mu}_1$
is the unique minimizer of the energy functional
\begin{equation}\label{squarerootproblem}
    I(\mu)+
    2\int_{0}^{\infty}\left(V(\sqrt{x})-a\sqrt{x}-U^{\mu_2}(\sqrt{x})\right) d\mu(x)
\end{equation}
over all probability measures $\mu$ on $[0,\infty)$.

We are going to show that the external field
$V(\sqrt{x})-a\sqrt{x}-U^{\mu_2}(\sqrt{x})$ in
\eqref{squarerootproblem} is convex for $x > 0$. Since
\[
U^{\mu_2}(\sqrt{x}) = -\frac{1}{2} \int \log(x+|z|^2)\, d\mu_2(z),
\]
we have
\[ \frac{d^2}{dx^2} \left( U^{\mu_2}(\sqrt{x}) \right) = \frac{1}{2} \int
    \frac{1}{(x+|z|^2)^2}\, d\mu_2(z),
\]
which due to the constraint $\mu_2\leq\sigma$ can be bounded by
\begin{align*}
    \frac{d^2}{dx^2} \left(  U^{\mu_2}(\sqrt{x}) \right) & < \frac{a}{2\pi} \int_{i\mathbb R}
    \frac{1}{(x+|z|^2)^2}\, |dz|  = \frac{a}{4x^{3/2}} \\
    & = \frac{d^2}{dx^2} \left(-a \sqrt{x} \right).
\end{align*}
It follows that $-a \sqrt{x} - U^{\mu_2}(\sqrt{x})$ is convex. Due to the
assumption in the lemma on the convexity of $V(\sqrt{x})$, it then follows that
$\left(V(\sqrt{x})-a\sqrt{x}-U^{\mu_2}(\sqrt{x})\right)$
 is indeed convex for $x > 0$.

From the convexity it follows that $\tilde{\mu}_1$ is supported on one interval.
Then $\mu_1$ is supported on either one or two intervals, depending on whether
$0$ belongs to the support of $\tilde{\mu}_1$ or not. The convexity also implies
that the density of $\tilde{\mu}_1$ does not vanish in the interior of its
support, and has square root behavior at its non-zero endpoint(s),  see e.g.\
\cite[Lemma 3.5]{CK}. The same properties then apply to $\mu_1$.
\end{proof}

The three cases in Section \ref{subsection:threecases} now specialize as
follows.

\paragraph{Case I:}
In the first case there are four real branch points $\pm b_1$, $\pm b_2$ (with
$b_1>b_2>0$) and no other branch points. The remaining zeros of the
discriminant \eqref{eq:D12} come as four double zeros. The Riemann surface has
genus zero.

\paragraph{Case II:}
In the second case there are four real branch points $\pm b_1$, $\pm b_2$ (with
$b_1>b_2>0$) and two purely imaginary branch points $\pm ic$ (with $c>0$).
The remaining zeros of the discriminant come in the form of a six-fold zero at
$z=0$, see Lemma \ref{lemma:caseII} below. The Riemann surface has genus one.

\paragraph{Case III:}
In the third case there are two real branch points at $\pm b_1$ (with $b_1>0$)
and two purely imaginary branch points at $\pm ic$ (with $c>0$). There are four
double zeros of the discriminant. The Riemann surface has genus zero.

\paragraph{} It remains to investigate in more detail the Cases I, II and III. More
precisely, for any $t\in\mathbb R$ and $a>0$ we want to determine which of the
three cases applies. We also want to find the curves in the $(t,a)$ plane where
a transition from one case to another takes place.

\subsection*{The genus one region}

Let us first investigate Case II. This is the case of genus 1. It turns out
that in this case, the parameters $\alpha$ and $\beta$ in the McLaughlin
equation \eqref{eq:McLbis} take on a particularly simple form: they are both equal to zero. This
is the content of the next lemma.

\begin{lemma} \label{lemma:caseII} (Case II.) Let $t\in\mathbb R$ and $a>0$
be such that the McLaughlin equation \eqref{eq:McLbis} is of genus 1. Then one
has that $\alpha=\beta=0$. Moreover, the discriminant of the McLaughlin
equation can then be factored as
\begin{equation}\label{eq:D126}
D_{12}(z) = z^6 D_6(z)
\end{equation} where $D_6$ is the degree six polynomial
\begin{multline}\label{eq:D6} D_6(z) = 4a^2
z^6+ \left(1-12ta^2\right) z^4 + \left((12t^2-18)a^2 - 2t\right) z^2\\
+\left(-27 a^4 +  (18t  -4t^3)a^2 - 4 + t^2\right).
\end{multline}
\end{lemma}

\begin{proof}
From the general descriptions above we know that in the genus 1 case, the
McLaughlin equation has six simple branch points $\pm b_1$, $\pm b_2$, and $\pm
ic$ ($b_1,b_2,c>0$), which are simple zeros of the discriminant. The remaining
zeros of the discriminant should come as three double zeros, possibly
coalescing. By symmetry $0$ is a double zero, and this forces $\alpha=0$, cf.\
\eqref{eq:D12}.

If we substitute $\alpha=0$ in \eqref{eq:McLbis} and calculate the discriminant
with respect to $\xi$ we obtain the $12$th degree polynomial
\begin{multline}\label{eq:D12:alpha=0}
4a^2z^{12}-(-4\beta+12a^2t-1)z^{10}-(18a^2-12t^2a^2+2t+12t\beta)z^8
\\-(4-12t^2\beta-t^2+18\beta-18a^2t+4t^3a^2+27a^4)z^6-(54a^2\beta-18t\beta+4t^3\beta)z^4-27\beta^2
z^2.
\end{multline}
We can take out the factor $z^2$ in \eqref{eq:D12:alpha=0}, corresponding to
the double zero at $z=0$. The remaining factor is of degree $10$ and we
consider it as a $5$th degree polynomial $D_5(y)$ in $y=z^2$. We already know
that this polynomial has three simple roots $b_1^2$, $b_2^2$, $-c^2$ and one
double root, call it $d$. Since by \eqref{eq:D12:alpha=0} the sum of the roots
equals $(-4\beta+12a^2t-1)/(4a^2)\in\mathbb R$, it follows that $d$ is real.
Moreover, from \eqref{eq:D12:alpha=0} we see that the product of the roots
equals $27\beta^2/(4a^2)\geq 0$. Since this product can also be written as
$-b_1^2b_2^2c^2d^2\leq 0$, it then follows that $\beta=d=0$.

In conclusion, we have shown now that $\alpha=\beta=0$. Inserting this in the
McLaughlin equation and computing its discriminant by a direct calculation then
leads to \eqref{eq:D126}--\eqref{eq:D6}.
\end{proof}

Note that the factor $z^6$ in \eqref{eq:D126} corresponds to the six-fold zero
at $z=0$, while the zeros of $D_6(z)$ should yield the branch points $\pm b_1$,
$\pm b_2$ and $\pm ic$. In particular, four of these zeros should be real and
the other two purely imaginary. The next lemma describes when this happens.

\begin{figure}[t]
\centering
\begin{overpic}[width=14cm]{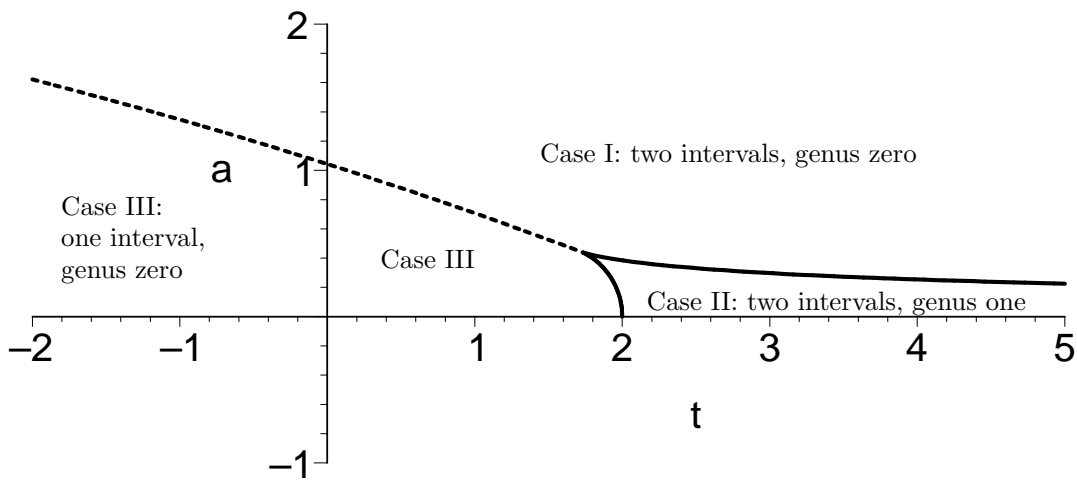}
    \put(50,30){Case I: two intervals, genus zero}
    \put(60,16){Case II: two intervals, genus one}
    \put(5,25){Case III:}
    \put(5,22){one interval,}
    \put(5,19){genus zero}
    \put(35,20){Case III}
\end{overpic}
\caption{Phase diagram for the quartic potential $\frac{1}{4}x^4 - \frac{t}{2} x^2$.
The dark curve represents the Painlev\'e II transition.
The dotted curve represents the Pearcey transition.}
\label{fig:PhaseDiagram}
\end{figure}

\begin{lemma}\label{lemma:genus1region}
The polynomial \eqref{eq:D6} has four real and two purely imaginary
zeros precisely for those $(t,a)\in\mathbb R\times \mathbb R_{+}$ lying in the
open region $\mathcal D$ bounded by the points $(t_1,a_1)=(2,0)$, $(t_2,a_2) =
(\sqrt{3},1/\sqrt[4]{27})$, $(t_3,a_3)=(+\infty,0)$, the straight line segment
between $(t_1,a_1)$ and $(t_3,a_3)$, and the two branches of the curve
\begin{equation}\label{eq:Painleve2curve}
D_6(0) = -27 a^4 +  (18t -4t^3)a^2 -
4 + t^2 = 0 \end{equation}
that connect the point $(t_2,a_2)$ with $(t_1,a_1)$
and $(t_3,a_3)$, respectively. The region $\mathcal{D}$ is shown in the bottom
right part of Figure~{\rm\ref{fig:PhaseDiagram}}.
\end{lemma}

\begin{proof} Rewrite $D_6(z)$ as a cubic polynomial in the variable $y=z^2$:
\begin{multline*} 4a^2 y^3 + \left(1-12ta^2\right) y^2 + \left((12t^2-18)a^2 - 2t\right) y \\
+\left(-27 a^4 +  (18t  -4t^3)a^2 - 4 + t^2\right).
\end{multline*}
We are looking for the values of $(t,a)\in\mathbb R\times\mathbb R_+$ for which
this cubic polynomial has two strictly positive and one strictly negative zero.
This is a standard routine whose description we omit. \end{proof}

The two previous lemmas show that the genus of the McLaughlin equation can only
be 1 if $(t,a)$ lies in the region $\mathcal D$. Outside $\mathcal D$ the genus
must necessarily be zero.

It remains to show that inside $\mathcal D$ the genus is exactly 1 (and not 0).
This is taken care of by the next lemma.
\begin{lemma}
\begin{itemize}
\item[\rm (a)]
Inside $\mathcal D$ the genus is either identically $1$ or identically $0$.
\item[\rm (b)] There exists at least one point $(t,a)$ in $\mathcal D$
for which the genus is $1$.
\end{itemize}
\end{lemma}

\begin{proof}
For Part (a) we invoke the following continuity result which follows from
general potential theory: The parameters $\alpha$ and $\beta$ in the McLaughlin
equation depend continuously on $t\in\mathbb R$ and $a>0$. Using this, one sees
that the discriminant of the McLaughlin equation also depends continuously on
$t$ and $a$ and therefore also the branch points since these are simple roots
of the discriminant.

Now let $\mathcal D_1\subset\mathcal D$ be the region formed by those
$(t,a)\in\mathcal D$ for which the genus is 1. We show that $\mathcal D_1$ is
both open and closed in $\mathcal D$. To show that it is open, let
$(t,a)\in\mathcal D_1$. Then the discriminant has six simple zeros and by
continuity the same holds in an open neighborhood of $(t,a)$. To show that
$\mathcal D_1$ is closed in $\mathcal D$, we take a sequence of points
$(t_k,a_k)\in\mathcal D_1$, $k=1,2,\ldots$, which converge to a limit point
$(t,a)\in\mathcal D$. By Lemma~\ref{lemma:caseII} we have $\alpha=\beta=0$ for
each $(t_k,a_k)$ so by continuity the same must hold for the limit point
$(t,a)$. But then Lemma~\ref{lemma:genus1region} shows that the discriminant
has six distinct simple zeros, which implies that the genus is 1. Hence
$(t,a)\in\mathcal D_1$.

For Part (b), we only outline a proof. The idea is to show that for any fixed
$t>2$, we have $(t,a)\in\mathcal D_1$ for all $a$ small enough. This relies on
the fact that for $t> 2$ the eigenvalues in the unitary matrix model with
potential $\frac{1}{4} x^4 - \frac{t}{2} x^2$ (without external source) are
supported on two intervals~\cite{BI1,BI2}. The claim then follows from a
continuity argument for $a\to 0$; we do not go into the details.

An alternative approach to prove Part (b) would be to pick a numerical point
$(t,a)\in\mathcal D$ and show by direct means (using the McLaughlin equation
with $\alpha=\beta=0$) that this algebraic curve makes the RH steepest descent
analysis work, in a similar vein as in \cite{ABK,BK2,BK3}.
\end{proof}

We summarize our findings with the following
\begin{proposition} Let $\mathcal D$ be the region defined in
Lemma~{\rm\ref{lemma:genus1region}}. For $(t,a)\in\mathcal D$ the McLaughlin
equation has genus one and we have $\alpha=\beta=0$. For $(t,a)\in(\mathbb
R\times\mathbb R_+)\setminus\mathcal D$ the genus is zero.
\end{proposition}

\subsection*{The genus zero region}

Now we focus on the genus zero region $(t,a)\in(\mathbb R\times\mathbb
R_+)\setminus\mathcal D$. A useful representation for the parameters $\alpha$
and $\beta$ in this region is given in \cite{ALT}. It is shown there that the
following parametrization holds
\begin{align}\label{eq:alpha}
    \alpha & = \frac{(1-u)(c^4+2c^4u-u^2)^2}{c^2u^3}\\
    \label{eq:beta}
    \beta & = -\frac{(3c^4-u)(c^4+2c^4u-u^2)^2}{c^4u^2}
\end{align}
where
\begin{equation}\label{eq:u} u:=2c^4+ac
\end{equation}
and where $c$ is a root of the equation
\begin{equation}\label{eq:c} 2c^6-2tc^4+3c^2-ac^3-tac-a^2 = 0.
\end{equation}
According to \cite{ALT}, $c$ is actually the \emph{largest positive} root of
equation \eqref{eq:c}, but we will not need this in what follows.


Let us seek the values of $(t,a)\in(\mathbb R\times\mathbb
R_+)\setminus\mathcal D$ for which a phase transition occurs between the Cases
I, II, III. For such $(t,a)$ the discriminant of the McLaughlin equation should
have a zero at $z=0$. By virtue of \eqref{eq:D12} this implies $\alpha=0$. From
\eqref{eq:alpha} this implies that either $u=1$, or $c^4+2c^4u-u^2=0$. Let us
first consider the case $u=1$. Then we have from \eqref{eq:u} that
\begin{equation}\label{eq:cbis} 2c^4+ac=1.
\end{equation} Thus the equations \eqref{eq:c}--\eqref{eq:cbis} have a common root $c$.
In other words, the resultant of these two equations with respect to the
variable $c$ should be zero. Computing this resultant with Maple yields the
following condition on $(t,a)$:
\begin{equation}\label{eq:Painleve2andPearcey} 54 a^4 + (72t - t^3) a^2 - (t^4 -16 t^2 + 64) = 0.
\end{equation}
Next we consider the case $c^4+2c^4u-u^2=0$. From
\eqref{eq:alpha}--\eqref{eq:beta} this implies that $\alpha=\beta=0$ and we
know from earlier considerations (or from a similar resultant calculation as
above) that this is only possible if $(t,a)$ is such that
\eqref{eq:Painleve2curve} holds. The phase transition on this curve will be
discussed in the next section. The phase transition on the curve
\eqref{eq:Painleve2andPearcey} will be discussed in the section thereafter.

\subsection*{Painlev\'e II transition}

At the two curved boundaries of the region $\mathcal D$ in
Figure~\ref{fig:PhaseDiagram}, we have a transition from genus 0 to genus 1.
Recall that these boundaries are described by the relevant branches of the
equation
\[ D_6(0) = -27 a^4 +  (18t  -4t^3)a^2 - 4 + t^2  = 0. \]
More precisely, these branches are given by
\begin{align*}
    A_1 = a^2 & = \frac{t}{3} - \frac{2}{27} \left[t^3 - (t^2-3)^{3/2}\right],\quad t\geq\sqrt{3} \\
    A_2 = a^2 & = \frac{t}{3} - \frac{2}{27} \left[t^3 + (t^2-3)^{3/2}\right],\quad \sqrt{3}\leq t\leq 2.
    \end{align*}
We have $A_1 > 0$ for every $t \geq \sqrt{3}$ whereas $A_2 > 0$ only for
 $\sqrt{3}\leq t< 2$ and $0 < A_2 \leq A_1$ for these $t$-values.
See Figure~\ref{fig:PhaseDiagram}.

On the above curves we have a transition from genus 0 to genus 1, and we expect
that the phase transition is of Painlev\'e~II type \cite{BI2,CK,DelKui1}. More
precisely, we expect that the following happens. If one lets $a$ decrease
towards the curve $a^2 = A_1$ (with $t>\sqrt{3}$), then the constraint $\sigma$
on the imaginary axis becomes active, and we have a transition from Case~I to
Case~II. If one further decreases $a$ towards the curve $a^2 = A_2$ ($\sqrt{3}<
t< 2$), then the gap in the support of $\mu_1$ closes and hence we have a
transition from Case~II to Case~III.

On the curve $a^2 = A_2$ the phase transition involves the eigenvalue measure
$\mu_1$ and therefore we expect Painlev\'e~II behavior in the local eigenvalue
correlations at the origin \cite{BI2,CK}. On the curve $a^2 = A_1$, however,
the phase transition takes place on the \lq non-physical\rq\ sheets of the
Riemann surface and therefore it is not felt in the eigenvalue statistics. But
then we expect Painlev\'e~II behavior in the recurrence coefficients for the
associated multiple orthogonal polynomials, as in~\cite{DelKui1}.

Note that for $t>2$ the transition at $a^2 = A_2$ does not occur. This is
consistent with the fact that for $t> 2$ the eigenvalues in the unitary matrix
model with potential $\frac{1}{4} x^4 - \frac{t}{2} x^2$ (without external
source) are supported on two intervals~\cite{BI1,BI2}, as mentioned before.

\subsection*{Pearcey transition}

Now we consider the equation \eqref{eq:Painleve2andPearcey},
\[ 54 a^4 + (72t - t^3) a^2 - (t^4 -16 t^2 + 64) = 0. \]
This equation has the two solutions
\begin{align*}
    A_3 = a^2 & = - \frac{2t}{3} + \frac{1}{108} \left[t^3 + (t^2+24)^{3/2} \right] \\
    A_4 = a^2 & = - \frac{2t}{3} + \frac{1}{108} \left[t^3 - (t^2+24)^{3/2}
    \right].
    \end{align*}
The branch $A_4$ is negative, and so is irrelevant for us. The other branch
$A_3$ is positive and we expect that for $t < \sqrt{3}$ a phase transition of
the Pearcey type \cite{AvM1,BK3,OR,TW1} takes place for $a^2 = A_3$. See
Figure~\ref{fig:PhaseDiagram}.

\section*{Acknowledgements}

The first author is supported in part by the National Science Foundation (NSF) Grant DMS-0652005.

The second author is a Postdoctoral Fellow of the Fund for Scientific Research -
Flanders (Belgium).

The third author is supported in part by FWO-Flanders project G.0427.09,
by K.U. Leuven research grant OT/08/33, by the Belgian Interuniversity Attraction Pole P06/02, by the
European Science Foundation Program MISGAM, and by grant
MTM2008-06689-C02-01 of the Spanish
Ministry of Science and Innovation.


\begin{thebibliography}{99}
\bibitem{AvM1}
    M. Adler and P. van Moerbeke,
    PDE's for the Gaussian ensemble with external source and the Pearcey
    distribution,
    Comm. Pure Appl. Math. 60 no. 9 (2007), 1261--1292.
\bibitem{ABK}
    A.I. Aptekarev, P.M. Bleher, and A.B.J. Kuijlaars,
    Large $n$ limit of Gaussian random matrices with external
    source, part II, Comm. Math. Phys. 259 (2005), 367--389.
\bibitem{ALT}
    A.I. Aptekarev, V.G. Lysov, and D.N. Tulyakov,
    Global eigenvalue distribution regime with an anharmonic potential and an external source,
    Theoretical and Mathematical Physics 159 (2009), 447--467.
\bibitem{BI1}
    P.M. Bleher and A. Its,
    Semiclassical asymptotics of orthogonal polynomials,
    Riemann-Hilbert problem, and universality in the matrix model,
    Ann. Math. 150 (1999), 185--266.
\bibitem{BI2}
    P.M. Bleher and A. Its,
    Double scaling limit in the random matrix model: the Riemann-Hilbert approach,
    Comm. Pure Appl. Math. 56 (2003), 433--516.
\bibitem{BI3}
    P.M. Bleher and A. Its,
    Asymptotics of the partition function of a random matrix
    model,
    Ann. Inst. Fourier 55 (2003), no. 6, 1943--2000.
\bibitem{BK1}
    P.M. Bleher and A.B.J. Kuijlaars,
    Random matrices with external source and multiple orthogonal polynomials,
    Int. Math. Research Notices 2004, no 3 (2004), 109--129.
\bibitem{BK2}
    P.M. Bleher and A.B.J. Kuijlaars,
    Large $n$ limit of Gaussian random matrices with external
    source, part I, Comm. Math. Phys. 252 (2004), 43--76.
\bibitem{BK3}
    P.M. Bleher and A.B.J. Kuijlaars,
    Large $n$ limit of Gaussian random matrices with external
    source, part III: double scaling limit,
    Comm. Math. Phys. 270 (2007), 481--517.
\bibitem{BH1}
    E. Br\'ezin and S. Hikami, Universal singularity at the closure
    of the gap in a random matrix theory, Phys. Rev. E 57 (1998),
    4140--4149.
\bibitem{BH2}
    E. Br\'ezin and S. Hikami, Level spacing of random matrices in
    an external source, Phys. Rev. E 58 (1998), 7176--7185.
\bibitem{CK}
    T. Claeys and A.B.J. Kuijlaars,
    Universality of the double scaling limit in random matrix
    models,
    Comm. Pure Appl. Math. 59 (2006), no. 11, 1573--1603.
\bibitem{CV}
    T. Claeys and M. Vanlessen,
    Universality of a double scaling limit near singular edge points in random matrix models,
    Comm. Math. Phys. 273 (2007), 499--532.
\bibitem{Dei}
    P. Deift, Orthogonal Polynomials and Random Matrices: a Riemann-Hilbert
    approach. Courant Lecture Notes in Mathematics Vol. 3, Amer. Math. Soc.,
    Providence R.I. 1999.
\bibitem{DKM}
    P. Deift, T. Kriecherbauer, and K.T-R McLaughlin,
    New results on the equilibrium measure for logarithmic potentials in the presence of an external
    field,
    J. Approx. Theory 95 (1998), 388--475.
\bibitem{DKMVZ1}
    P. Deift, T. Kriecherbauer, K.T-R McLaughlin, S. Venakides, and  X. Zhou,
    Uniform asymptotics for polynomials orthogonal with respect to
    varying exponential weights and applications to universality
    questions in random matrix theory,
    Comm. Pure Appl. Math. 52 (1999), 1335--1425.
\bibitem{DKMVZ2}
    P. Deift,  T. Kriecherbauer, K.T-R  McLaughlin, S. Venakides, and  X. Zhou,
    Strong asymptotics of orthogonal polynomials with respect to exponential weights,
    Comm. Pure Appl. Math. 52 (1999), 1491--1552.
\bibitem{DelKui1}
    S. Delvaux and A.B.J. Kuijlaars,
    A phase transition for non-intersecting Brownian motions, and the Painlev\'e
    II equation, Int. Math. Res. Not. (2009), 3639--3725.
\bibitem{Dubr}
    B. Dubrovin,
    Theta functions and non-linear equations,
    Russian Math. Surveys 36 (1981), 11--92.
\bibitem{DuK}
    M. Duits and A.B.J. Kuijlaars,
    Universality in the two matrix model: A Riemann-Hilbert steepest descent
    analysis, Comm. Pure Appl. Math. 62 (2009), 1076--1153.
\bibitem{DuKuMo}
    M. Duits, A.B.J. Kuijlaars and M.Y. Mo,
    The Hermitian two matrix model with an even quartic potential,
    in preparation.
\bibitem{Dyson}
    F.J. Dyson,
    A Brownian-motion model for the eigenvalues of a random matrix,
    J. Math. Phys. 3 (1962), 1191--1198.
\bibitem{FK}
    H. M. Farkas and I. Kra,
    Riemann Surfaces.
    Graduate Texts in Mathematics Vol. 71, Springer-Verlag, New York--Berlin. 1980.
\bibitem{Fay}
    J. Fay,
    Theta Functions on Riemann Surfaces.
    Springer-Verlag, Berlin, 1973.
\bibitem{FIK}
    A.S. Fokas, A.R. Its, and A.V. Kitaev,
    The isomonodromy approach to matrix models in 2D quantum gravity,
    Commun. Math. Phys. 147 (1992), 395--430.
\bibitem{Ku2}
    A.B.J. Kuijlaars,
    Multiple orthogonal polynomial ensembles,
    in: Recent Trends in Orthogonal Polynomials and Approximation Theory
    (Arves\'u et al., eds.), Contemp. Math. Vol. 507, Amer. Math. Soc., Providence, R.I., 2010,
    preprint arXiv:0902.1058.
\bibitem{KDragnev1}
    A.B.J. Kuijlaars and P.D. Dragnev, Equilibrium problems associated with fast
    decreasing polynomials, Proc. Amer. Math. Soc. 127 (1999), 1065–-1074.
\bibitem{KMFW}
    A.B.J. Kuijlaars, A. Mart\'{\i}nez-Finkelshtein and F. Wielonsky,
    Non-intersecting squared Bessel paths and multiple orthogonal polynomials for modified Bessel weights,
    Comm. Math. Phys. 286 (2009), 217--275.
\bibitem{KMo}
    A.B.J. Kuijlaars and M.Y. Mo,
    The global parametrix in the Riemann-Hilbert steepest descent analysis for
    orthogonal polynomials, preprint arXiv:0909.5626.
\bibitem{McL}
    K. T-R McLaughlin,
    Asymptotic analysis of random matrices with external source and a family of algebraic curves,
    Nonlinearity 20 (2007), 1547--1571.
\bibitem{Mo}
    M.Y. Mo,
    Universality in the two matrix model with a monomial quartic and a general
    even polynomial potential.
    Comm. Math. Phys. 291 (2009), 863--894.
\bibitem{NS}
    E.M. Nikishin and V.N. Sorokin,
    Rational Approximations and Orthogonality,
    Amer. Math. Soc., Providence, RI, 1991.
\bibitem{OR}
    A. Okounkov and N. Reshetikhin,
    Random skew plane partitions and the Pearcey process,
    Comm. Math. Phys. 269 (2007), 571--609.
\bibitem{Pastur}
    L. Pastur,
    The spectrum of random matrices (Russian),
    Teoret. Mat. Fiz. 10 (1972), 102-112.
\bibitem{SaffTotik}
    E.B. Saff and V. Totik,
    Logarithmic Potentials with External Field,
    Springer-Verlag, Berlin, 1997.
\bibitem{TW1}
    C. Tracy and H. Widom,
    The Pearcey process,
    Comm. Math. Phys. 263 (2006), 381--400.
\bibitem{VAGK} W. Van Assche, J.S. Geronimo and A.B.J. Kuijlaars,
    Riemann-Hilbert problems for multiple orthogonal polynomials,
     Special Functions 2000: Current Perspectives and Future Directions
    (J. Bustoz et al., eds.), Kluwer, Dordrecht, 2001, pp. 23--59.
\bibitem{ZJ1}
    P. Zinn-Justin,
    Random Hermitian matrices in an external field,
    Nuclear Phys. B, 497 (1998), 725-–732.
\bibitem{ZJ2}
    P. Zinn-Justin,
    Universality of correlation functions of Hermitian random
    matrices in an external field,
    Comm. Math. Phys. 194 (1998), 631–-650.
\end{thebibliography}
\end{document}